\documentclass[aps,pre,twocolumn,amsmath,amssymb,nofootinbib,floatfix]{revtex4-1}

\usepackage{amsmath}
\usepackage{amssymb}
\usepackage{amsthm}
\usepackage[dvipdf]{color}
\usepackage{graphicx}
\usepackage{dcolumn}
\usepackage{bm,subfigure} 
\usepackage{xcolor,colortbl}

\DeclareMathOperator{\im}{Im}

\DeclareMathOperator{\tr}{tr}

\DeclareMathOperator{\rank}{rank}

\begin{document}
\title{Universal Continuum Theory for Topological Edge Soft Modes}

  \author{Kai Sun}
  \author{Xiaoming Mao}

 \affiliation{
 Department of Physics,
  University of Michigan, Ann Arbor, 
 MI 48109-1040, USA
 }

\begin{abstract} 
Topological edge zero modes and states of self stress have been intensively studied in discrete lattices at the Maxwell point, offering robust properties concerning surface and interface stiffness and stress focusing.  In this paper we present a continuum topological elasticity theory and generalize the Maxwell-Calladine index theorem to continuous  elastic media. This theory not only serves as a macroscopic description for topological Maxwell lattices, but also generalizes this physics to continuous media. We define a \emph{gauge-invariant} bulk topological index, independent of microscopic details such as  choice of the unit cell and lattice connectivity. This index directly predicts the number of zero modes on edges, and it naturally extends to media that deviate from the Maxwell point, depicting how topological zero modes turn into topological soft modes. 
\end{abstract}

\maketitle

\emph{Introduction}---Topologically-protected edge zero modes (ZMs)  and states of self stress (SSSs) have been intensively studied recently in lattice models at the verge of mechanical instability~\cite{Kane2014,Lubensky2015,mao2018maxwell}, i.e., Maxwell lattices, leading to a rich collection of interesting phenomena including topological solitons~\cite{Chen2014}, mechanical Weyl modes~\cite{Rocklin2016,Stenull2016}, switchable surface stiffness~\cite{Rocklin2017}, programmable buckling~\cite{Paulose2015a} and fracturing processes~\cite{Zhang2018},  and beyond~\cite{Zhou2018,Zhou2019,Chen2015,Paulose2015,Suesstrunk2016,Socolar_2017,Rocklin_2017,Saremi2018}.

These topological phenomena are all based on discrete lattice and network models, and the physical principles that govern them are dependent of microscopic details, e.g., the 
connectivity of the networks.
However,  topological edge ZMs and SSSs are found to exist at long wave length (e.g., as opposed to isolated points in the Brillouin zone), where microscopic details become unimportant, which indicates that these topological phenomena
may not be exclusive for lattice models. Instead, they may also arise in continuum elasticity.  Preliminary continuum elasticity analysis using linear~\cite{Lubensky2015} and nonlinear strain tensor~\cite{Rocklin2017} identified two different regimes based on the intrinsic soft strain of the lattice: a dilation dominant regime where FMs localize on all edges with inverse decay length $\sim q_x$ (where $q_x$ is the wavevector parallel to the edge), and a shear dominant regime where low frequency ($\omega\sim q^2$ where $q$ is the wavevector) soft modes exist in the bulk. 
To obtain topological edge modes, which decay as $\kappa \sim O(q_x^2)$, and their topological index, lattice theory with optical bands was necessary.  
It remains an open question whether a universal  description independent of microscopic mechanisms for such topological phenomena exists.

Remarkably, it is shown recently that these topological edge modes still survive when the system deviates from the idealized discrete lattice limit, in specimen obtained from 3D printing~\cite{bilal2017intrinsically}, cutting~\cite{Ma2018}, and finite-element~\cite{Ma2018,ma2019influence} and discrete lattice~\cite{Stenull2019} analysis.  The original definition of the topological polarization $R_T$, based on the Maxwell condition, is no longer valid in these systems.  However edge soft modes similar to topological Maxwell lattices are observed, e.g., asymmetric edge stiffness and edge waves. Are these systems truly topological?

In this paper we present a universal continuum theory that can be directly applied to both continuous and discrete elastic media and predict topological soft modes and SSSs.  
We extend the ``Maxwell condition'' of the verge of mechanical instability from lattice models, where it takes the form of $\langle z \rangle=2d$ 
with $\langle z \rangle$ being the mean coordination number and $d$ being the spatial dimension)~\cite{Jacobs1995,Liu2010,Mao2010,Ellenbroek2011,Broedersz2011,Sun2012,Mao2015,Zhang2015a,Zhang2016}, to the continuum in terms of macroscopic elastic moduli.
We name elastic systems that satisfy this macroscopic  Maxwell condition  ``Maxwell media'', which include many Maxwell lattices and beyond.
We further define a universal way to measure how far a system is away from this macroscopic Maxwell condition. For
systems at or not far away from the Maxwell condition, a topological index $N$ can be defined, which predicts the number of ZMs for each edge. Remarkably, this topological index 
is insensitive to any microscopic details such as choice of the unit cell and is determined purely by macroscopic elastic constants.  
We also develop methods to obtain the elastic constants in the continuum theory from first principles, 
and verify our theory through good agreement with results on lattice models at and close to the Maxwell point.

\noindent{\it Elastic energy with higher-order gradients---}Within linear elasticity, the elastic energy is a quadratic form of displacement-field gradients~\cite{Landau1986}. 
Typically, only the first-order gradients are included, while higher-order derivative terms only contribute to short-distance physics~\cite{Mindlin1964} and thus 
can be ignored at long distance. 
However, as mentioned above, for Maxwell systems, the leading-order theory is insufficient, and 
thus a more generic setup with higher-order gradients becomes essential even for the long-distance limit.
As proved in the SI (Sec.~S-1),
the elastic energy of any stress-free elastic medium must take the following form
\begin{align}
E
=\int \mathbf{dr} \sum_{i=1}^{d(d+1)/2} \frac{\lambda_{i}}{2} &\left(\mathcal{M}^{(0)}_{ijk}\partial_{j}u_{k}+
\mathcal{M}^{(1)}_{ijk\ell}\partial_{j}\partial_{k}u_{\ell} +\ldots\right)^2
\nonumber\\
&+O(\partial\partial u)^2 ,
\label{eq:elastic:general}
\end{align}
where $u$ is the displacement field. 
This sum contains $d(d+1)/2$ terms where $d$ is the spatial dimension. 
In  standard elasticity theory (using the Voigt notation)~\cite{kittel1976introduction}, 
elastic moduli form a $[d(d+1)/2]\times [d(d+1)/2]$ symmetric matrix. Once this matrix is diagonalized, it gives  the leading order
terms of Eq.~\eqref{eq:elastic:general}, with $\lambda_i$ and $\mathcal{M}^{(0)}$ being the eigenvalues and eigenvectors of the Voigt elastic moduli matrix.
For topological indices, it is often sufficient to just include one higher order term into the theory, $\mathcal{M}^{(1)}$.

In addition to these $d(d+1)/2$ squares, the elastic energy also contains some higher-order terms [$O(\partial\partial\mathbf{u})^2$ and above], which will be ignored for now. 
These ignored terms are non-negative, and thus for ZMs (e.g., in Maxwell lattices), they must vanish. 
For a general elastic medium, they lead to subleading corrections to the elastic energy [$\propto O(q^4)$], which 
will be calculated below via a variational method.

\noindent{\it Maxwell's counting in continuum---}In continuum elasticity, the number of degrees of freedom is $d$ for each point in real space, because $u$ contains $d$ components. 
For each positive $\lambda_i>0$, the corresponding square term in Eq.~\eqref{eq:elastic:general} enforces one constraint at this point
\begin{align}
\mathcal{C}_{ij}u_{j}=0 \;\;\; \textrm{for} \;\;\; \lambda_i >0 ,
\label{eq:constraints:real:space}
\end{align}
where $\mathcal{C}_{ij}=\mathcal{M}^{(0)}_{ikj}\partial_{k}+\mathcal{M}^{(1)}_{ik\ell j}\partial_{k}\partial_{\ell} +\ldots$
If a deformation field satisfies all the constraints, it is a ZM. 

For a rigid solid,  all $\lambda_i$'s are positive and thus the constraint number is $d(d+1)/2$, which always exceeds $d$ for $d>1$,
i.e., the medium is over constrained.
This conclusion agrees with the generalized counting argument for continuous media~\cite{sun2019maxwell}, and 
 the Janet-Cartan embedding theorem~\cite{janet1926possibilite,cartan1927possibilite}.
For a system to be at the verge of mechanical instability, some $\lambda_i$ may vanish. Most importantly, if the matrix has $d$ positive eigenvalues,
while the rest are $0$, the number of constraints coincides with that of the degrees of freedom, which will be referred to as the macroscopic Maxwell condition,
and this type of systems will be called \emph{Maxwell media}.
In contrast to Maxwell lattices/networks, Maxwell media only require the Maxwell condition at the macroscopic level.
In a Maxwell medium the $\mathcal{C}$ matrix defined above becomes a square matrix and this matrix serves the role of the compatibility matrix  in Maxwell lattices.

\noindent{\it Zero modes and states of self stress---}Let us consider generic elastic media without assuming the Maxwell condition. The elastic energy [Eq.~\eqref{eq:elastic:general}]
can be written as
\begin{align}
E=\frac{1}{2}\int \mathbf{dr}\; \mathbf{u}^T\overleftarrow{\mathcal{C}}^T \Lambda \overrightarrow{\mathcal{C}}\mathbf{u} ,
\label{eq:energy:M:matrix}
\end{align}
where $\Lambda$ is the diagonal matrix composed of positive $\lambda_i$'s. Because $\Lambda$ is 
positive definite by definition, Eq.~\eqref{eq:energy:M:matrix} implies that any ZMs must satisfy $\mathcal{C}\mathbf{u}=0$.
It should be emphasized that in Eq.~\eqref{eq:energy:M:matrix}, derivatives in $\mathcal{C}$ act to the right on $\mathbf{u}$, while $\mathcal{C}^T$ to the left on $\mathbf{u}^T$, as marked by the arrows above. 
This differs  from the dynamical matrix $\mathcal{D}=- \overrightarrow{\mathcal{C}}_{\partial\to-\partial}^T \Lambda \overrightarrow{\mathcal{C}}$, where all derivatives act 
to the right. In $\mathcal{D}$, derivatives in $\mathcal{C}^T$ are flipped from left to right via integral by parts, which introduces boundary terms.
These boundary terms are irrelevant for bulk properties but crucial for physics at the edge. 
In particular, one needs to distinguish two conditions (a) $\mathcal{C}\mathbf{u}=0$ and (b) $\mathcal{D}\mathbf{u}=0$. 
From Eq.~\eqref{eq:energy:M:matrix}, it is easy to check that (a) is the sufficient and necessary condition of ZMs (i.e., $E=0$). 
In contrast, (b) is the sufficient and necessary condition for the system to be force free, because  force in an elastic medium is 
$f_i=-\delta E/\delta u_i=D_{ij}u_j$. The condition (b) is weaker than  (a), because 
force-free deformations can be classified into two categories: (1) ZMs ($\mathcal{C}\mathbf{u}=0$) 
and (2) modes that leave the bulk in force balance but there is force on the  boundary  ($\mathcal{D}\mathbf{u}=0$ but $\mathcal{C}\mathbf{u}\ne 0$).
Furthermore, similar to the lattice theory, $\mathcal{C}^T$ plays the role of the equilibrium  matrix and its null space contains SSSs of the medium, which yields all possible ways the medium can carry residual stress.  Interestingly, the Maxwell-Calladine index theorem that relates the number of ZMs and SSSs still holds  [see the SI (Sec.~S-1)].

\begin{figure}[t]
	\centering
	\subfigure[]{\includegraphics[width=.32\columnwidth]{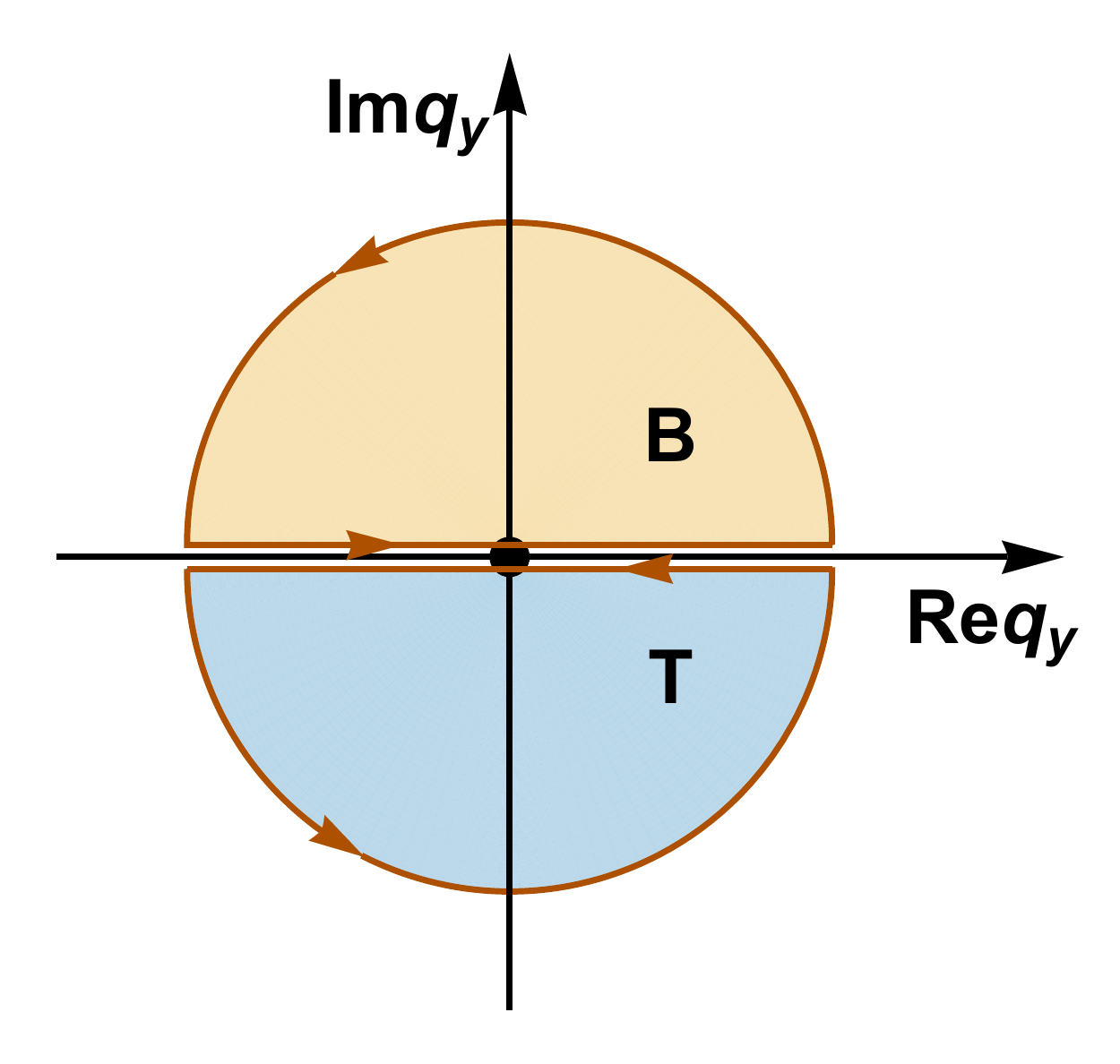}}
	\subfigure[]{\includegraphics[width=.32\columnwidth]{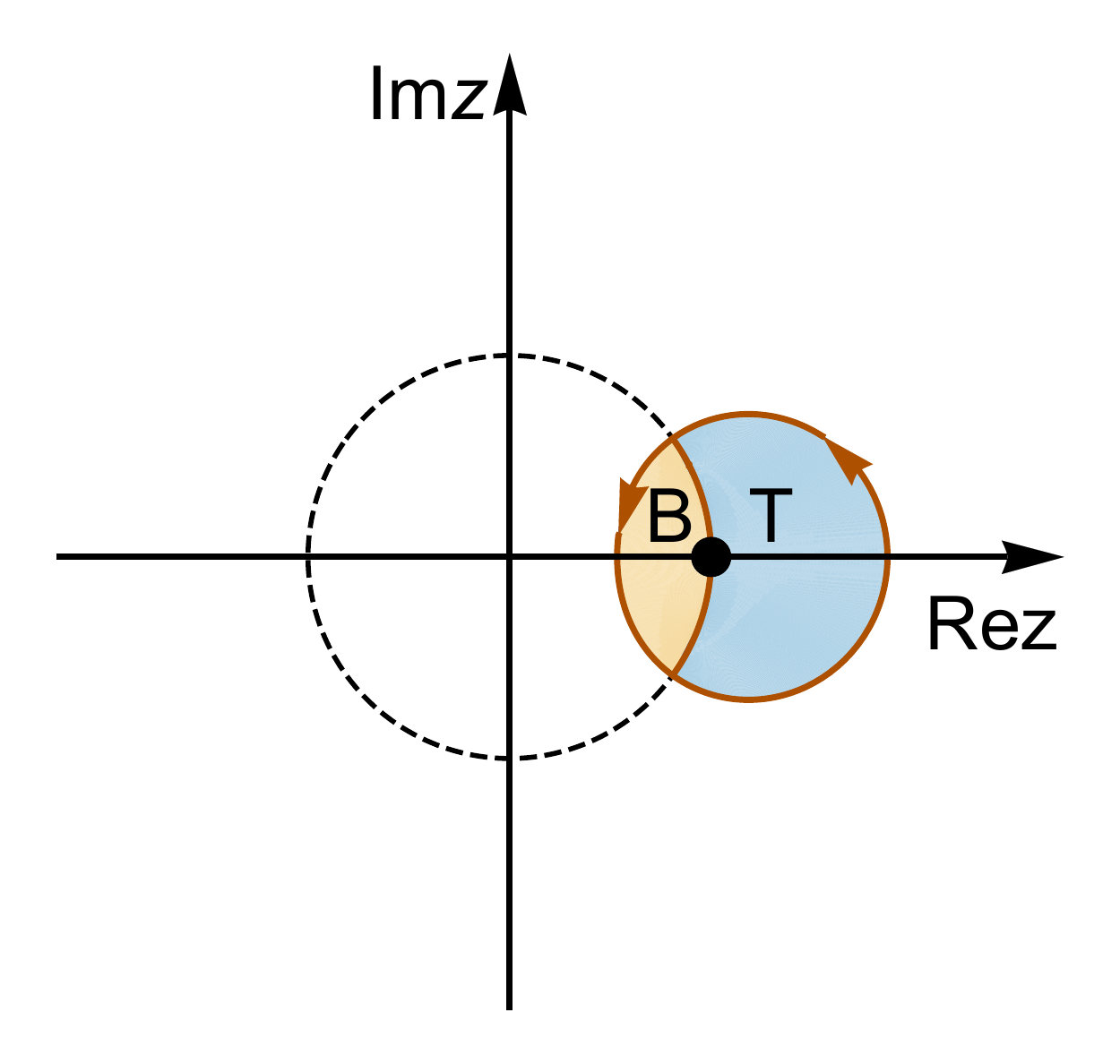}}
	\subfigure[]{\includegraphics[width=.32\columnwidth]{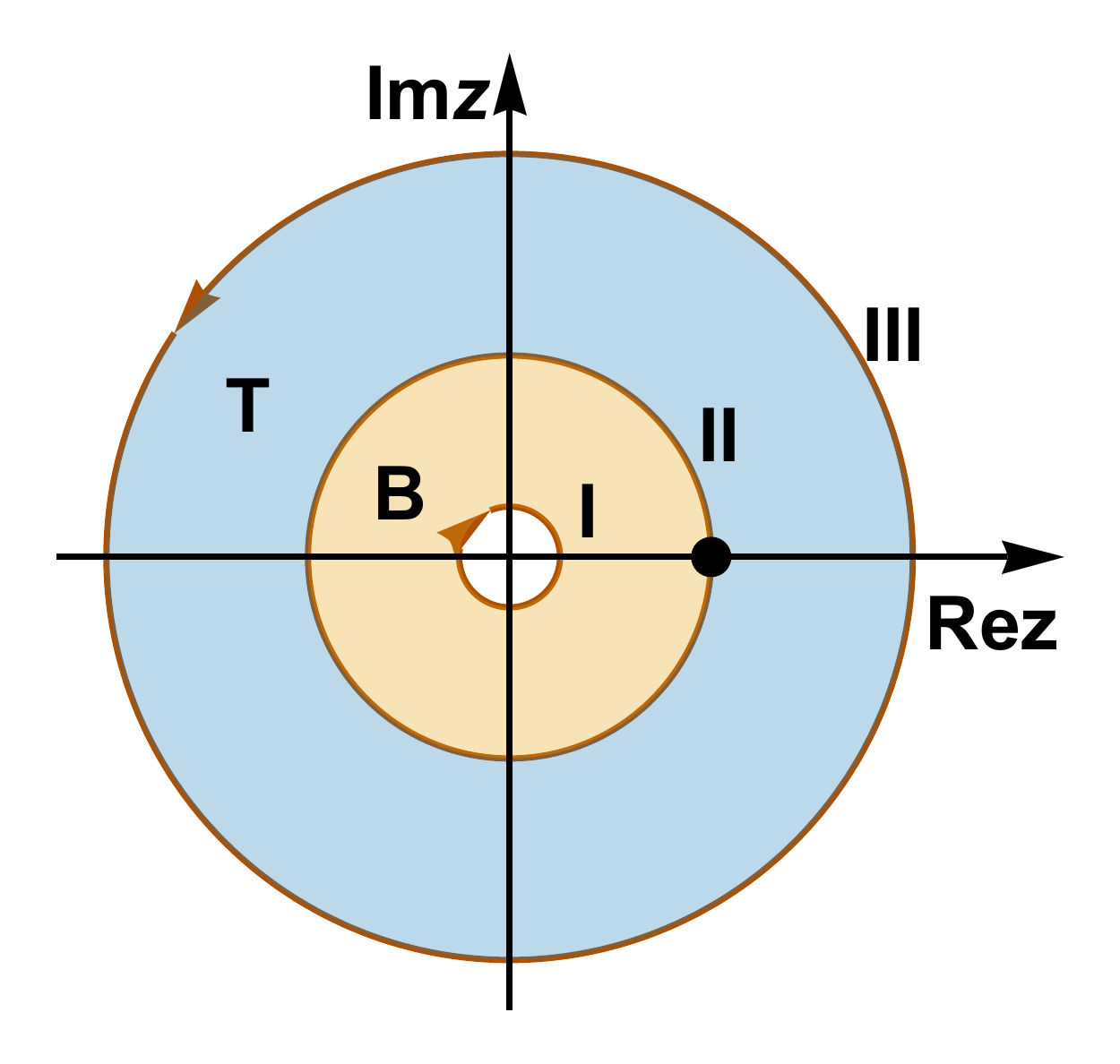}}
	\caption{Topological index integral contours in the complex (a) $q_y$ and  (b-c) $z$ plane, where $z= \exp(i q_y)$. 
	The contour around region $T$ ($B$) gives the topological index $N_T$ ($N_B$) for the top (bottom) edge. 
	The continuum theory captures long-wavelength physics near $q=0$ ($z=1$) marked by the black dot. The radius of the circle in (a) is $\Lambda_q$. 
	In discrete Maxwell lattices, the contour is usually taken as the unit circle in the z-plane, i.e., II in (c). In comparison, (b) shows the contour of (a) in terms of $z$.
	For lattice models, gauge-invariant topological indices can be obtained using contours shown in (c), i.e., the boundary of region B (contours I and II) or T (II and III).}
	\label{fig:contour}
\end{figure}

\noindent{\it The topological index---}Similar to lattices, edge ZMs in a Maxwell medium also come from nontrivial topology in the bulk and in this section we define the topological index, which
dictates the number of ZMs on each edge. We demonstrate the physics in a 2D system, which can be easily generalized to other spatial dimensions.

Consider an infinitely long strip parallel to the $x$ axis with two open edges (top/bottom).  As shown in the SI (Sec.~S-3), for such a Maxwell system, zero-energy edge states 
may arise, with deformation fields $\mathbf{u}=\mathbf{A} e^{i q_x x+ i q_y y}$, where $q_x$ is real and $q_y$ is complex, whose imaginary part
describes how fast the amplitude decays  into the bulk.
For each fixed $q_x=\bar{q}_x$, these ZMs are characterized by a bulk topological index.  
\begin{align}\label{eq:NTB}
N_{T/B}(\bar{q}_x)=\frac{1}{2\pi i}\oint_{T/B} d q_y \tr \left(\mathcal{C}^{-1} \partial_{q_y }\mathcal{C}\right) ,
\end{align}
where $\mathcal{C}_{ij}=i\mathcal{M}^{(0)}_{ikj}q_{k}-\mathcal{M}^{(1)}_{ik\ell j}q_{k}q_{\ell} +\ldots$
is the Fourier transform of the real-space $\mathcal{C}$ matrix in Eq.~\eqref{eq:constraints:real:space}.
For any closed contour, this index is quantized to integer values,   because it is the winding number of the phase of $\det \mathcal{C}$.
Here, the contour is a semi circle on the complex $q_y$ plane as shown 
in Fig.~\ref{fig:contour}(a).
If the semicircle is in the upper (lower) half of the plane, the integral 
defines $N_T$ ($N_B$), which gives the number of edge modes for the top (bottom) edge at $q_x=\bar{q}_x$. 
Similar to Maxwell lattices, these two indices characterize the topological structure of the bulk bands and their values remain invariant, unless the system undergoes 
a topological phase transition, which closes and the reopens the bulk phonon gap at $q_x=\bar{q}_x$.

The radius of the semicircle is the ultraviolet momentum cutoff $\Lambda_q$,
which arises in (almost) any continuum theory, separating universal properties at long distance ($q<\Lambda_q$) from microscopic details 
at short distance ($q>\Lambda_q$).
Although a cutoff is needed for its definition, this topological index is cutoff independent, as long as $1/\Lambda_q$
is much shorter (longer) than any macroscopic (microscopic) length scales. In our theory, the microscopic length scale can be estimated
as  $\mathcal{M}^{(1)}/\mathcal{M}^{(0)}$, where $\mathcal{M}^{(0)}$ and $\mathcal{M}^{(1)}$ are the characteristic elastic constants.

\noindent{\it Gauge invariance---}It is worthwhile to emphasize that the topological index defined above differs slightly from the topological polarization defined for 
Maxwell lattices, due to differences in the integral contours.
In Maxwell lattices, it is known that the topological polarization is gauge sensitive, e.g., by choosing a different unit cell,
its value may change by an integer.  Such a constant shift doesn't change the topological classification. However, as one tries to connect topological polarization 
with physical observables, e.g., the number of ZMs at an edge, the gauge dependent part needs to be carefully deducted. 
In Maxwell lattices, this is usually achieved utilizing a local polarization, a microscopic quantity that depends on local lattice connectivity at 
the edge~\cite{Kane2014,Lubensky2015,mao2018maxwell}.
For a continuum theory which doesn't rely on microscopic details, a local gauge-fixing is unavailable. Thus a universal 
gauge-fixing scheme independent of microscopic details becomes necessary. Remarkably, the integral contour defined above offers a  gauge independent topological index,
which directly gives the number of edge states for each edge.

This can be seen by comparing the contours of the continuum~[Eq.~\eqref{eq:NTB}] and the lattice theory.  In Maxwell lattices, because $q_y$ has a periodicity associated with the reciprocal lattice vector (the first Brillouin zone), 
 the integral contour is usually chosen to be along the real $q_y$ axis over one period (e.g., from $q_y=-\pi$ to $q_y=+\pi$). A change of variable $z=e^{i q_y}$ is then defined, 
 under which the contour becomes the unit circle in the complex $z$ plane, and components of the topological polarization %$R_T=\sum_i n_i a_i$ 
 is given by 
$n_i= \oint_i \frac{d z}{2\pi i}\ \tr (\mathcal{C}^{-1} \partial_{z }\mathcal{C})$.
As shown in the SI (Sec.~S-3), $\tr (\mathcal{C}^{-1} \partial_{z }\mathcal{C})$ is analytic for the entire complex $z$ plane, except for a finite number of poles.
Thus the integral above only comes from residue at each pole, and there are two types of poles.  
Each ZM contributes one pole with residue $1$, and their contributions are gauge invariant.  This is the reason why $R_T$ can probe ZMs. 
However, there is a second type of poles at $z=0$ and $z=\infty$.
 Residue at these two poles are gauge dependent.
These two poles are  not related with physical ZMs, and they are natural consequences of analytic continuation.  Thus, their contribution needs to be removed.  
In Maxwell lattices this is done via the local polarization $R_L$.

In this continuum theory, this gauge fixing is automatically achieved, 
because these two points are beyond the ultraviolet cutoff [Fig.~\ref{fig:contour}(b)]. This  scheme can also be applied to Maxwell lattices, 
utilizing integral contours  in Fig.~\ref{fig:contour}(c).

It needs to be emphasized that this gauge-invariant topological index does not detect edge modes 
with zero penetration depth, such as a dangling bond. This is because zero penetration depth 
corresponds to $|\im q_y|=\infty$ (i.e., $z=0$ or $z=\infty$), which has been removed in our definition. 
Because such a zero-penetration-depth edge mode is purely due to local and microscopic features at the edge, a bulk topological index is not expected to detect them. 
Instead, they can only be detected via microscopic local probes, such as a local polarization~\cite{Kane2014,Lubensky2015,mao2018maxwell}.

\noindent{\it Optical phonon modes---}In certain soft materials, such as the regular kagome lattice~\cite{Sun2012}, low-energy optical modes may exist, whose energy can become comparable with acoustic modes.
% For example, in a kagome lattice, the optical phonon mode associated with rotating the triangles can reach zero frequency~\cite{Sun2012}.  
These soft optical modes can be included in the continuum theory to more accurately capture the low-energy physics, especially at intermediate $q$.
With soft optical modes, the elastic energy becomes
\begin{align}
E%
=\int \mathbf{dr} &\sum_i  \frac{\lambda_{i}}{2} \big( \mathcal{M}^{(0)}_{ijk}\partial_{j}u_{k} 
+ \mathcal{M}^{(0)}_{ij}\phi_j+\mathcal{M}^{(1)}_{ijk\ell}\partial_{j}\partial_{k}u_{\ell}
\nonumber\\
&
+\mathcal{M}^{(1)}_{ijk}\partial_{j}\phi_{k}
+\ldots\big)^2
+O[(\partial\partial u)^2,(\partial \phi)^2] ,
\label{eq:elastic:general:optical}
\end{align}
where $\phi_i$  represents the soft optical modes, and the $m$th order derivatives of $\mathbf{u}$ should be treated as the same order as the $(m-1)$th order derivatives of $\phi_i$. 
The number of degrees of freedom is now increased to $d+n_O$ at each point in real space, with $n_O$ representing the number of soft optical modes included.
The number of eigenvalues ($\lambda_i$) is now $d(d+1)/2+n_O$. Same as before, if all $\lambda_i$'s are positive, the system is over constrained (for $d>1$), and 
when $d(d-1)/2$ of the eigenvalues vanishes, the Maxwell condition is achieved, at which the topological indices and all the physics can be easily generalized.
The same construction [Eq.~\eqref{eq:elastic:general:optical}] also applies to pinned systems, 
e.g., the pinned 1D chain model discussed in Ref.~\cite{Kane2014}, where all phonon modes are optical-like, and $\mathbf{u}$ should be removed. 
%because of the absence of acoustic modes.

\begin{figure}[t]
	\centering
	\subfigure[]{\includegraphics[width=.48\columnwidth]{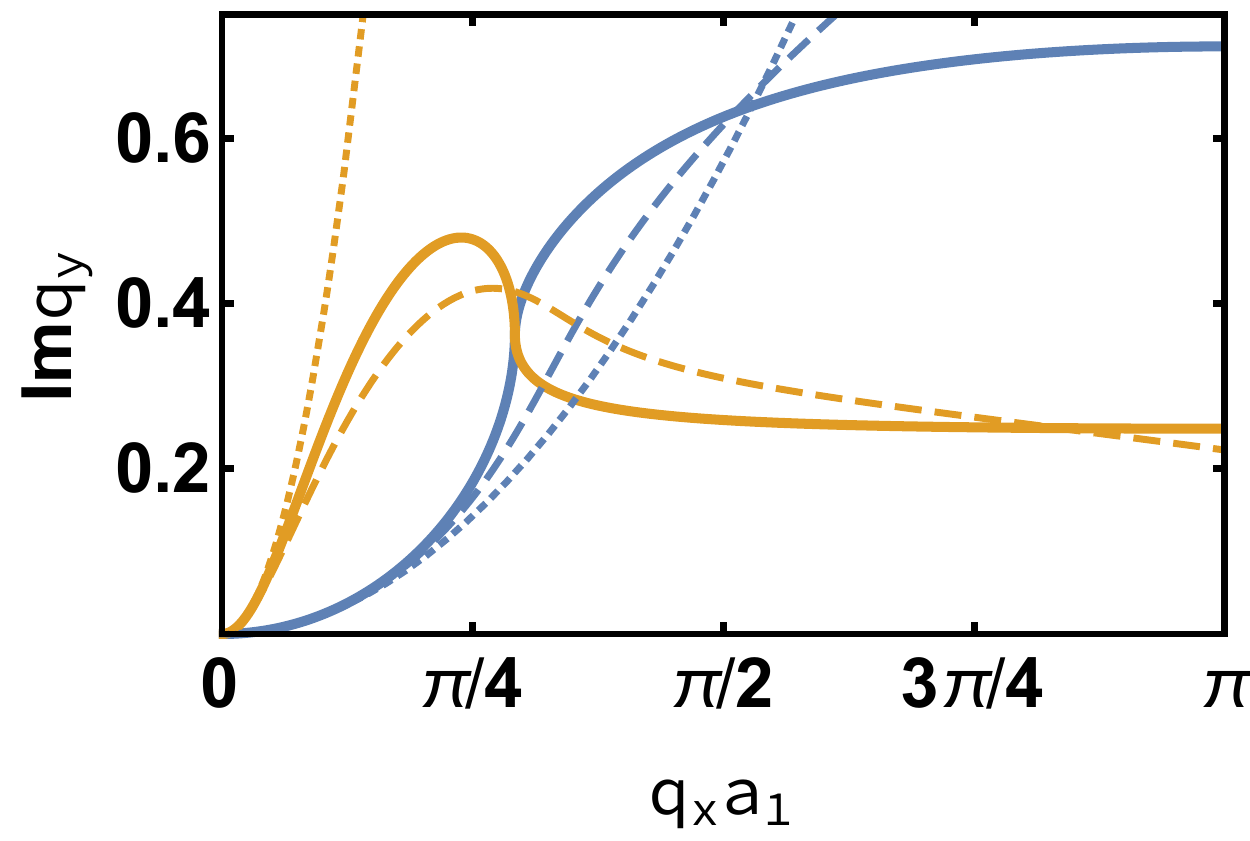}}
	\subfigure[]{\includegraphics[width=.48\columnwidth]{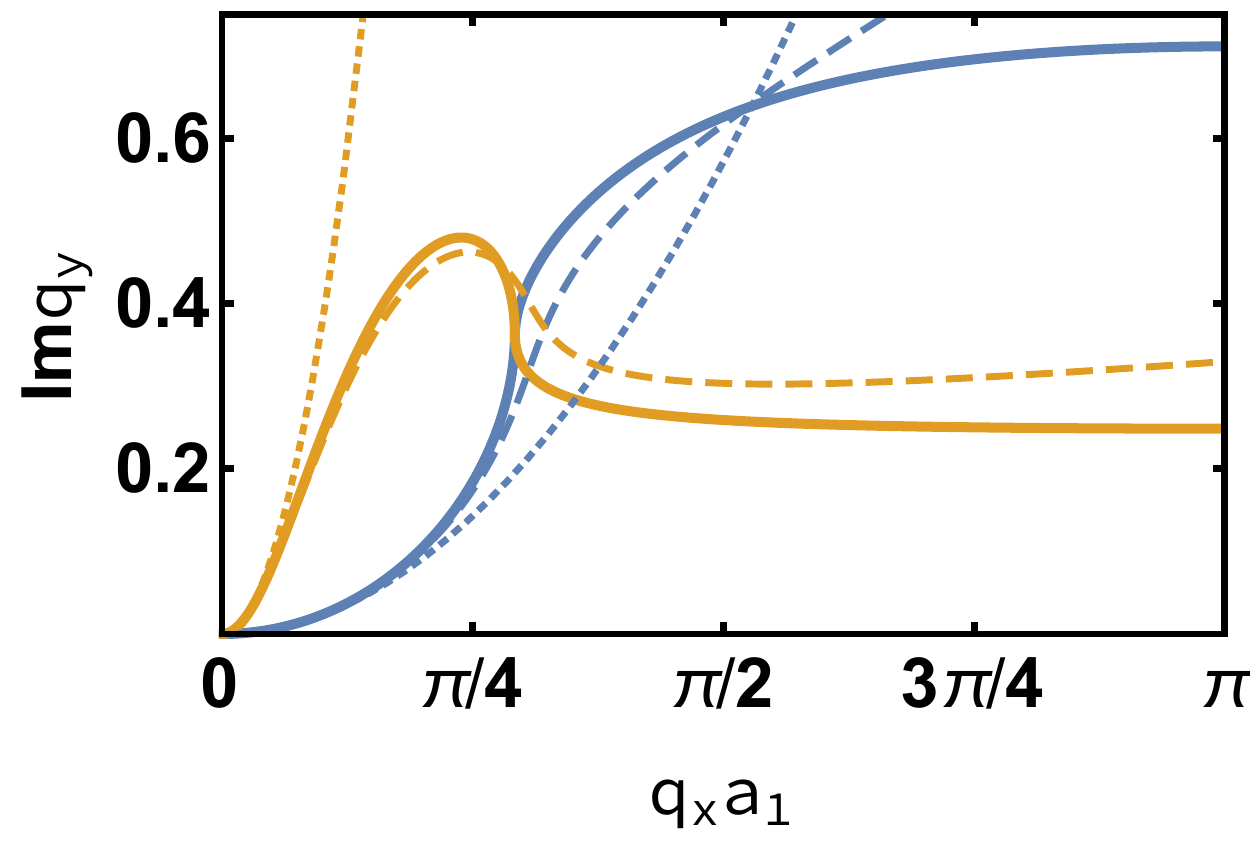}}
	\caption{Decay rates of topological edge modes in a topological kagome lattice [defined in the SI (Sec.~S-5)]. 
	$q_x$ and $a_1$ are the wavevector and the lattice constant along the open edge.
	Solid lines show the decay rates of two edge modes computed from the lattice theory, while the dashed lines are
	obtained from the continuum theory, whose small-$q$ asymptotic forms are shown as dotted lines. In (a), the continuum theory contains two acoustic modes, while (b)
	includes one more soft optical mode.}.
	\label{fig:decay:rate}
\end{figure}

\noindent{\it Comparison with Maxwell lattices---}In the SI (Sec.~S-2), we present a first-principle method to determine the coefficients in our continuum theory, based on lattice models or finite-element method (FEM). The key idea in this method
is to integrate out high-energy optical modes, which are irrelevant in the long-distance low-energy continuum theory. For the dynamic matrix, this is a standard practice~\cite{Mao2011a}. However,
because the topological information is stored in the $\mathcal{C}$ matrix instead, here we need to develop a new approach based on the $\mathcal{C}$ matrix.
Remarkably, it can be proved that as far as ZMs are concerned, our procedure of integrating out high-energy modes is an exact method, 
although the number of degrees of freedom is  reduced. Next, we then take the long wavelenghth limit, which leads to the continuum theory.

With this first-principle technique, we can now directly compare the continuum theory with lattice models or FEM.
In Fig.~\ref{fig:decay:rate}, we show the decay rates of the edge modes in a topological kagome lattice and compare it with the continuum theory [SI (Sec.~S-5)].
Fig.~\ref{fig:decay:rate}(a) shows the continuum theory with only acoustic modes [Eq.~\eqref{eq:elastic:general}], while (b) contains one extra soft optical mode 
[Eq.~\eqref{eq:elastic:general:optical}]. Both of them provide the correct topological index, fully agreeing with the lattice model, and the decay rate from the continuum theory is 
asymptotically exact at long wavelength. Remarkably, the continuum theory
even captures fine details at intermediate $q_x$. Near $\pi/4$, the two edge modes happen to become degenerate with the same $q_y$, which is also seen in
the continuum theory. The two continuum models in (a) and (b) share identical asymptotic behavior at small $q_x$, while 
including the soft optical branch improves the accuracy at intermediate $q_x$,
where energy of the soft optical mode becomes comparable with the acoustic ones.

\noindent{\it Maxwell medium beyond Maxwell lattices---}For the deformed kagome lattice, the continuum theory predicts two edge ZMs, agreeing with the microscopic Maxwell counting ~\cite{Sun2012, Kane2014,Lubensky2015,mao2018maxwell}.
For this lattice, the microscopic counting guarantees that the energy of these modes remain zero, even when higher-order gradients beyond  Eq.~\eqref{eq:elastic:general} are included.
However, for a generic Maxwell medium, terms ignored in Eq.~\eqref{eq:elastic:general} [i.e., $O(\partial\partial u)^2 $ terms] may push the energy of the edge modes up (i.e., they become soft modes instead of ZMs), even into the bulk bands. 
As a result, for these systems, in general, after using Eq.~\eqref{eq:elastic:general} to obtain the topological index and edge modes, 
we need to put these edge-mode candidiates back into the full elastic energy to check whether their energy is indeed below bulk phonon bands.
Only if the energy is below the bulk bands, the candidiate becomes as a real edge soft mode. 

As shown in the SI (Sec.~S-4), this calculation can be easily implemented via a variational method. We utilize each edge-mode candidiate obtained from the topological 
theory as a trial wavefunction and minimize the  full elastic energy. As a variational method, this minimization is exact for ZMs but tends to 
 overestimate the energy 
of  finite-frequency modes ~\cite{griffiths2004introduction}. Thus if it reveals an edge soft mode with energy lower than the bulk band, 
the accurate energy of this mode must be even lower, and thus it must be indeed an edge mode.

\begin{figure}[t]
	\centering
	\subfigure[]{\includegraphics[width=.48\columnwidth]{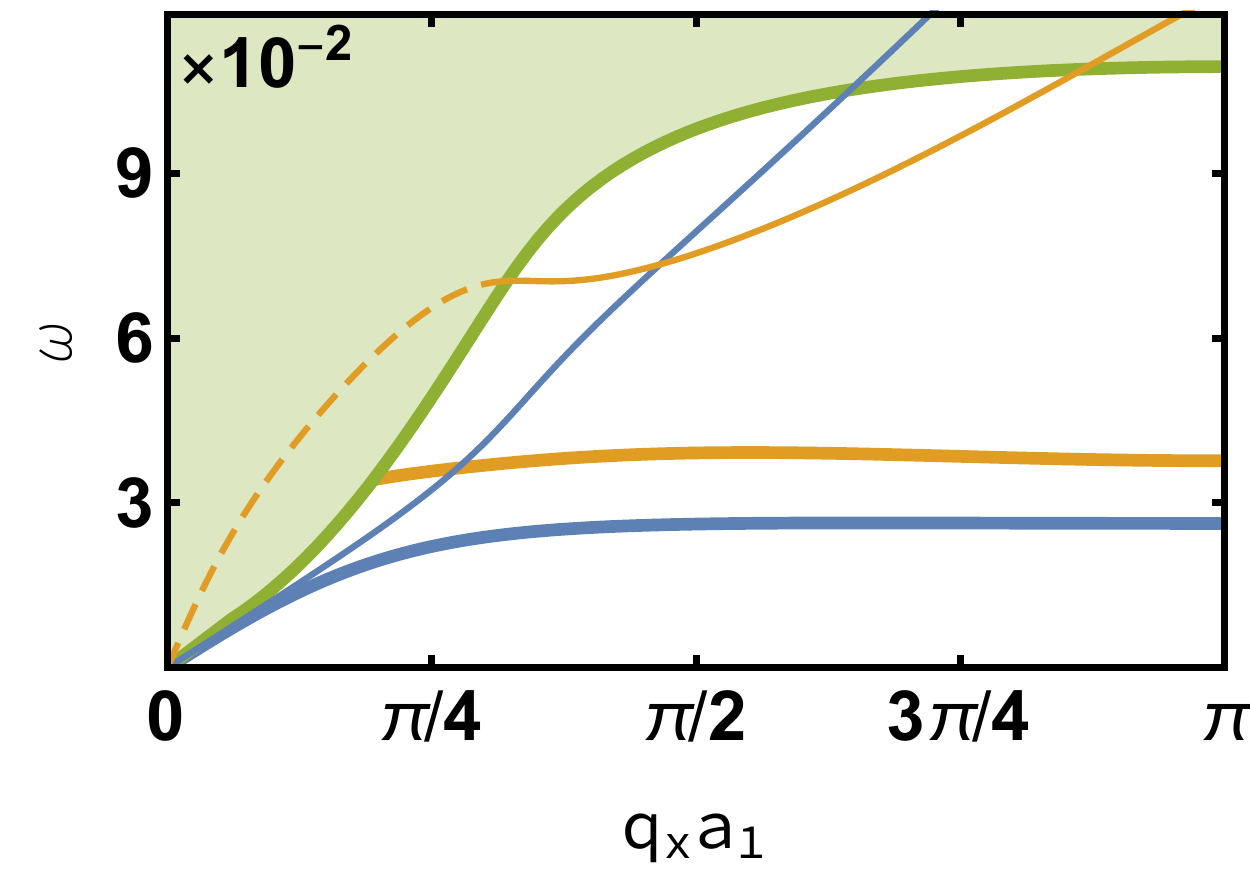}}
	\subfigure[]{\includegraphics[width=.48\columnwidth]{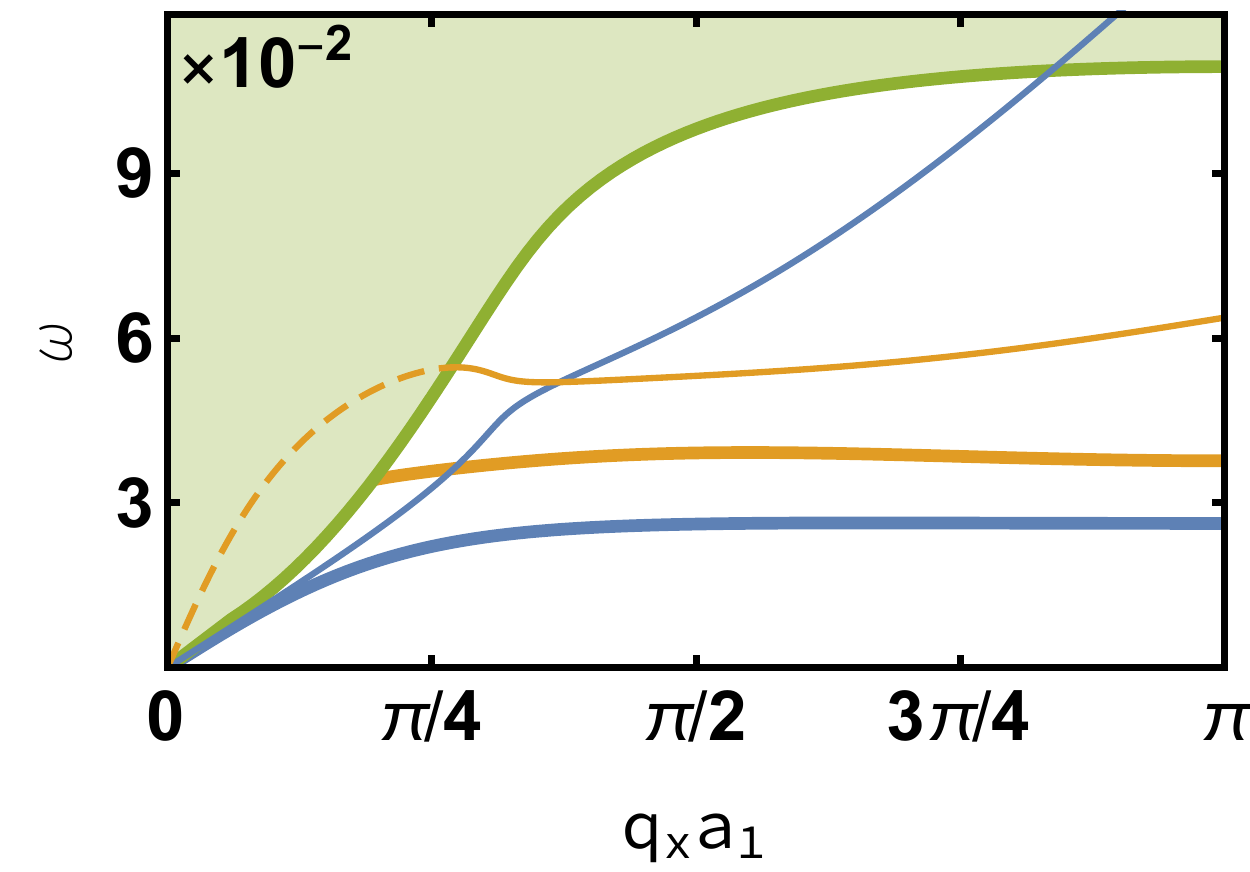}}
	\caption{Edge modes in an over-constrained kagome lattice [defined in the SI (Sec.~S-5)].
	The shaded area marks the energy continuum from bulk bands. Thick (thin) solid lines are the dispersion of the two edge modes on the soft edge from lattice model
	(continuum theory with first-order energy correction). $r_M$ in (a) and (b) are $0.066125$ and $0.0369656$ respectively.}
	\label{fig:energy}
\end{figure}

\noindent{\it Beyond Maxwell medium---} Maxwell media require some of the $\lambda_i$'s to be exactly $0$. 
In a real elastic solid  all $\lambda_i$'s are finite and positive. 
Here, we can order the eigenvalues from high to low and compute the ratio between the $(d+n_O+1)$th largest eigenvalue and the $(d+n_O)$th, $r_M=\lambda_{d+n_O+1}/\lambda_{d+n_O}$, where $n_O$ is the number of soft optical modes included in the continuum theory.
$r_M$  offers a universal measurement to determine how far away a real system is from the ideal Maxwell limit. 
Its value is between $0$ and $1$,  while $0$ implies a Maxwell medium.
If this quantity is small enough, one can set all $\lambda_{i}$'s with $i>d+n_O$ to zero in the leading-order treatment, and perform the topological analysis as shown above.
Then, similar to the previous section, we compute the energy of edge-mode candidates using the variational method.

In Fig.~\ref{fig:energy}, we study a deformed kagome lattice
with next-nearest-neighbor springs, whose spring constant is 1/1000 times the nearest-neighbor ones. This system is over constrained, 
but is not far away from the Maxwell point. Thus, it exhibits similar edge modes. We compute the phonon band structure on an infinitely-long 
strip with two open edges ($200$ unit cells wide), and compare it with  predictions from the continuum theory and the variational method,
and find good agreement. 
The continuum theory used in  Fig.~\ref{fig:energy}(a) includes only acoustic modes, while (b) contains one extra optical mode.
It must be emphasized that in contrast to the lattice numerical solution, which is performed on a strip with open edges, curves from the continuum theory are 
computed in the  wave vector space without any physical edge, but  this wave vector space treatment correctly captures the long-distance physics at an open edge. 

\noindent{\it Acknowledgements---}  
We thank S. Gonella for enlightening discussions which inspired this work.  This work was supported by the National Science Foundation under Grant No. NSF-EFRI-
1741618.

\begin{widetext}
\section*{\Large Supplementary Information}
\section{Generic elastic theory}
In this section, we discuss the elastic theory of a generic elastic medium. We will stay in the linear response regime. However, higher-order derivative terms of the deformation field 
are allowed and will be included, which play a crucial role in the topological description. The key  result of this section is that 
for any stress-free elastic medium, the elastic energy can always be organized 
in the following form
\begin{align}
E%
=\frac{1}{2}\int \mathbf{dr} \sum_{i=1}^{d(d+1)/2} \bigg[
\lambda_{i}\left(\mathcal{M}^{(0)}_{ijk}\partial_{j}u_{k}+\mathcal{M}^{(1)}_{ijk\ell}\partial_{j}\partial_{k}u_{\ell} 
+\mathcal{M}^{(2)}_{ijk\ell m }\partial_{j}\partial_{k}\partial_{\ell}u_m 
+\ldots\right)^2
\nonumber\\
+ \sum_{i=1}^{d^2(d+1)/2} \widetilde{\lambda}_{i} \left(\widetilde{\mathcal{M}}^{(1)}_{ijk\ell}\partial_{j}\partial_{k}u_{\ell} +\widetilde{\mathcal{M}}^{(2)}_{ijk\ell m }\partial_{j}\partial_{k}\partial_{\ell}u_m  +\ldots \right)^2
\nonumber\\
+ \sum_{i=1}^{(d+2)(d+1)d^2/6}\widetilde{\widetilde{\lambda}}_{i} \left(\widetilde{\widetilde{\mathcal{M}}}^{(2)}_{ijk\ell m }\partial_{j}\partial_{k}\partial_{\ell}u_m  +\ldots \right)^2
\nonumber\\
+\ldots \bigg] ,
\label{app:eq:elastic:full}
\end{align}
where $u$ is the displacement field (i.e., each point moves $r \to r+u(r)$), and $i,j,k,\ldots$ are the Cartesian indices.  
Here, we rewrite the elastic energy as sets of square terms with non-negative coefficients ($\lambda_i\ge 0$, $\widetilde{\lambda}_i\ge 0$, 
$\widetilde{\widetilde{\lambda}}_i\ge 0$, $\ldots$).
The first set contains $d(d+1)/2$ terms, each of which is the square of some linear combination
of deformation-field gradients. The second sets contains $d^2(d+1)/2$ squares. Here, each square term is also composed of a linear combination of deformation-field gradients,
but the derivatives  starts from the second order (i.e., no first-order derivative terms). The third set contains $(d+2)(d+1)d^2/6$ terms, and the linear combination starts from 
third-order gradient terms, etc.

\subsection{A simple example}
\label{app:sec:simple:exmpale}
We first use a simple example to demonstrate the key ideas.
Consider a real quadratic form $f(x,y)=a x^2 +2 b x y+ c y^2$, where $a$, $b$ and $c$ are three real constants, and $x$ and $y$ are two real variables.
Here, we show here that if the quadratic form is \emph{semi-positive definite}, i.e. $f(x,y)\ge 0$ for any real $x$ and $y$, 
then it can always be written as the sum of two squares 
\begin{align}
f(x,y)=a (x+d y)^2+ e y^2
\label{app:eq:quadratic:from}
\end{align}
with $a\ge 0$ and $e\ge 0$. 

For a semi-positive definite quadratic form, $a$ must be non-negative, i.e., $a\ge 0$.
If $a>0$, Eq.~\eqref{app:eq:quadratic:from} can be achieved by absorbing the cross term $b x y$ into $x^2$, which gives $d=b/a$ and $e=c-b^2/a$.
If $a=0$, the $x^2$ term vanishes and thus, we couldn't use it to absorb the $x y$ term. However, for a semi-positive-definite quadratic form, 
it is easy to verify that if $a=0$, $b$ must also vanish. Thus, when $a=b=0$, $f(x,y)= c y^2$ and
Eq.~\eqref{app:eq:quadratic:from} automatically holds as long as we set $e=c$.

It must be emphasized that Eq.\eqref{app:eq:quadratic:from} only applies to semi-positive-definite quadratic forms. This is very important, because it sets an important
constraint to our theory. As will be shown below, our theory only applies to stress-free elastic media, because stress-free is the necessary and sufficient condition
to ensure that the quadratic form used in our theory is semi-positive definite.

\subsection{Elastic energy with higher order derivative terms}
In this section, we use $\mathbf{u}$ to represent the displacement field. All the $m$th-order gradients of the displacement field will be made into a column vector, which will be dubbed
as $\mathbf{u}^{(m)}$. Here, the first-order terms are jus the linearized strain tensor (presented in the Voigt notation), e.g., in 2D $\mathbf{u}^{(1)}=(u_{xx},u_{yy},u_{xy})^T$
where $T$ stands for transpose.
 For higher order terms with $m>1$, $\mathbf{u}^{(m)}$ contains 
%$d \begin{pmatrix}
%i+d-1 \\
%d-1
%\end{pmatrix}$ 
$d (d+m-1)!/[(d-1)!m!]$ components, where $d$ is the spatial dimension and $!$ represents the factorial, e.g., in 2D,
$\mathbf{u}^{(2)}=(\partial_{x}^2 u_x, \partial_{x}\partial_{y} u_x, \partial_{y}^2 u_x, 
\partial_{x}^2 u_y, \partial_{x}\partial_{y} u_y, \partial_{y}^2 u_y)$. 

As will be proved in the next section, for any stress-free system, the elastic energy (within linear elasticity) must be a quadratic form of these $\mathbf{u}^{m}$'s,
\begin{align}
E=\frac{1}{2}\int \mathbf{dr}\;
\begin{pmatrix}
\mathbf{u}^{(1)} & \mathbf{u}^{(2)} &  \mathbf{u}^{(3)} & \ldots
\end{pmatrix}
\begin{pmatrix}
\mathcal{K}^{(1,1)} & \mathcal{K}^{(1,2)} & \mathcal{K}^{(1,3)} & \ldots \\
\mathcal{K}^{(2,1)} & \mathcal{K}^{(2,2)} & \mathcal{K}^{(2,3)} &  \ldots \\
\mathcal{K}^{(3,1)} & \mathcal{K}^{(3,2)} & \mathcal{K}^{(3,3)} & \ldots \\
\vdots & \vdots & \vdots & \ddots \\
\end{pmatrix}
%\mathcal{K}
\begin{pmatrix}
\mathbf{u}^{(1)} \\ \mathbf{u}^{(2)} \\ \mathbf{u}^{(3)}  \\ \vdots
\end{pmatrix}
\label{app:eq:elastic:energy:block:matrix}
\end{align}
The matrix in Eq.~\eqref{app:eq:elastic:energy:block:matrix}, which will be dubbed the $\mathcal{K}$ matrix is a \emph{semi-positive-definite symmetric real} matrix, 
composed of elastic constants.

Below we  prove that a semi-positive definite $\mathcal{K}$  can be decomposed into the following form
$\mathcal{K}=\mathcal{L} \mathcal{D}\mathcal{L}^T$
\begin{align}
\begin{pmatrix}
\mathcal{K}^{(1,1)} & \mathcal{K}^{(1,2)} & \mathcal{K}^{(1,3)} & \ldots \\
\mathcal{K}^{(2,1)} & \mathcal{K}^{(2,2)} & \mathcal{K}^{(2,3)} &  \ldots \\
\mathcal{K}^{(3,1)} & \mathcal{K}^{(3,2)} & \mathcal{K}^{(3,3)} & \ldots \\
\vdots & \vdots & \vdots & \ddots \\
\end{pmatrix}=
\begin{pmatrix}
\mathcal{L}^{(1,1)}) & 0 & 0 & \ldots \\
\mathcal{L}^{(2,1)}) & \mathcal{L}^{(2,2)} & 0 &\ldots \\
\mathcal{L}^{(3,1)}) & \mathcal{L}^{(3,2)} & \mathcal{L}^{(3,3)} &\ldots \\
\vdots & \vdots & \vdots & \ddots \\
\end{pmatrix}
\begin{pmatrix}
\mathcal{D}^{(1,1)} & 0 & 0 & \ldots \\
0 & \mathcal{D}^{(2,2)} & 0&\ldots \\
0 & 0 & \mathcal{D}^{(3,3)} &\ldots \\
\vdots & \vdots & \vdots & \ddots \\
\end{pmatrix}
\begin{pmatrix}
\mathcal{L}^{(1,1)} & 0 & 0 & \ldots \\
\mathcal{L}^{(2,1)} & \mathcal{L}^{(2,2)} & 0 &\ldots \\
\mathcal{L}^{(3,1)} & \mathcal{L}^{(3,2)} & \mathcal{L}^{(3,3)} &\ldots \\
\vdots & \vdots & \vdots & \ddots \\
\end{pmatrix}^T
%\begin{pmatrix}
%(\mathcal{L}^{(1,1)})^T & (\mathcal{L}^{(2,1)})^T & (\mathcal{L}^{(3,1)})^T & \ldots \\
%0 & (\mathcal{L}^{(2,2)})^T & (\mathcal{L}^{(2,3)})^T &\ldots \\
%0 & 0 & (\mathcal{L}^{(3,3)})^T &\ldots \\
%\vdots & \vdots & \vdots & \ddots \\
%\end{pmatrix}
\end{align}
This decomposition is a block version of the so-called LDLT decomposition, where the matrix $\mathcal{D}$ in the middle is a diagonal matrix. $\mathcal{L}$ is a block 
lower-triangular matrix and  $\mathcal{L}^T$ is its transpose. After this decomposition, it is easy to verify that the elastic energy takes the form
of Eq.~\eqref{app:eq:elastic:full}.

\subsection{Stress-free systems and the $\mathcal{K}$ matrix}
In this section, we prove that the elastic energy of a stress-free system must take the form of Eq.~\eqref{app:eq:elastic:energy:block:matrix}
and the $\mathcal{K}$ matrix must be semi-positive definite. For the leading order theory (i.e., $u^{(1)}$ only), this is obvious. However, because higher-order derivatives of the displacement field are included here, this conclusion needs to be proved with this more generic setup. In particular, because higher-order gradient terms can be 
obtained from lower-order ones (by taking derivatives), globally, $u^{(m)}$ and $u^{(n)}$ (with $m\ne n$) are not independent, and thus a careful proof becomes necessary.

Here, we define \emph{stress-free} based on stability against cutting. We cut a system into smaller pieces. If every piece can maintain its original shape after cutting,
no matter how we cut it,  the system is called stress-free. With this definition, it is easy to check that for a stress-free system, 
its elastic energy needs to be \emph{locally} semi-positive definite, i.e., for any deformation, the elastic energy density at every single point needs to increase or stay the same.
This is in contrast to a stressed system, whose elastic energy only needs to be \emph{globally} semi-positive definite, i.e., for any deformation, the total elastic energy
must increase or stay the same, but the elastic energy in some local region could decrease. Obviously, if we cut this local region out from the system, it  cannot maintain its shape.
Instead, it will deform to further reduce the elastic energy.

Early on, we mentioned that globally, $u^{(m)}$ and $u^{(n)}$ (with $m\ne n$) are related, because we can get higher-order derivatives terms
from lower-order ones by taking derivates. Here, we prove that $u^{(m)}$'s are locally independent, i.e.,
in a local region, we can design the displacement field to make $u^{(m)}$'s into any values that we want. This can be achieved via the Taylor series.
Without loss of generically, here we consider a local region around the origin. Here, we can define a displacement field in the form of a Taylor series
\begin{align}
u_i(x,y)= a_{ix} x + a_{iy} y + \frac{1}{2}(a_{ixx} x^2 + 2a_{ixy} x y+ a_{iyy} y^2)+\ldots
\end{align}
Near the origin, coefficients in this Taylor series are just the values of $u^{(m)}$'s, and thus by selecting the values of these coefficients, we can achieve arbitrary $u^{(m)}$'s.
It must be emphasized that this is a local construction and is only valid in a domain near the origin. As we move far away from the origin, this deformation may hit
singularities, at which the series stops to converge.

The local construction is already enough for us, because as shown above for a stress-free system, the elastic energy density needs to be semi-positive definite for 
any local regions in the system. This immediately implies that for any points in the system, the elastic energy density cannot contain linear terms of $u^{(m)}$'s. 
If some linear $u^{(m)}$ terms exists, we can always find some $u^{(m)}$'s that make the energy density at this point negative, and as shown above,
such a set of $u^{(m)}$'s  can always be realized using the construction above. This means that if we cut this local region out, it will deform to lower its energy, and thus the system
is stressed. This proves that for a stress-free system, 
the elastic energy density must start from quadratic order of $u^{(m)}$'s. If we ignore higher order terms (i.e., in the linear response regime),
this is a quadratic form as shown in Eq.~\eqref{app:eq:elastic:energy:block:matrix}.

Following the same line of thinking, we can prove that the quadratic form must be semi-positive definite in a stress-free system. If the $\mathcal{K}$ matrix contains negative 
eigenvalues, then we can design the stress field as shown above to make $u^{(m)}$'s the eigenvector of this negative eigenvalue in this local region. This will make the local
energy density negative, which is prohibited for stress free systems.

In summary, for a stress-free system, $\mathcal{K}$ is semi-positive definite. Thus, we can use the idea demonstrated in Sec.~\ref{app:sec:simple:exmpale} 
to absorb cross terms into squares.

\subsection{Leading-order theory}
At the leading order (i.e., $\mathbf{u}^{(1)}$ only), Eq.~\eqref{app:eq:elastic:energy:block:matrix} recovers the standard elastic theory presented in the Voigt notation.
In the Voigt notation, the stress tensor is written as a $d_c=d(d+1)$ vector, $\mathbf{u}^{(1)}$, where $d_c=d(d+1)/2$ is known as the Cartan dimension~\cite{cartan1927possibilite}.
The elastic energy is a quadratic form of the stress tensor
\begin{align}
E=\frac{1}{2}\int \mathbf{dr}\;\mathbf{u}^{(1)} 
\mathcal{K}^{(1,1)}
\mathbf{u}^{(1)} 
=\frac{1}{2} \int \mathbf{dr} \;  \sum_{i=1}^{d_c}\lambda_{i} \left(\mathcal{O}_{ij}u^{(1)}_j\right)^2
\label{eq:elastic:leading}
\end{align}
Here, $\mathcal{K}^{(1,1)}$ is the $d_c\times d_c$ real symmetric matrix, which contains $d_c(d_c+1)/2$ elastic moduli, 
among which $d(d-1)/2$ of them can be made zero by choosing proper coordinate orientation~\cite{Landau1986}.
As shown in Eq.~\eqref{eq:elastic:leading}, via diagonalizing the $\mathcal{K}^{(1,1)}$ matrix, the elastic energy can be written as the sum of $d_c=d(d+1)$ terms, 
each of which is the square of certain linear combination of $\mathbf{u}^{(1)}$. Here, $\mathcal{O}$ is an orthogonal matrix, which diagonalizes $\mathcal{K}^{(1,1)}$ 
[$\mathcal{O} \mathcal{K}^{(1,1)} \mathcal{O}^T = \textrm{diag}(\lambda_1, \lambda_2, \ldots \lambda_{d_c})$ with $\lambda_{i}$ 
representing the eigenvalues of $\mathcal{K}^{(1,1)}$].
For stability reasons, $\mathcal{K}^{(1,1)}$  must be semi-positive definite  and thus the eigenvalues $\lambda_i$ must be non-negative. 
Following Maxwell's counting argument, for each positive eigenvalue $\lambda_i>0$, the corresponding square term
enforces one constraint
\begin{align}
\mathcal{O}_{ij} u^{(1)}_j=\mathcal{O}_{ij} \mathcal{P}_{jk \ell} \partial_k u_\ell=0  \;\;\; \textrm{for} \;\;\; \lambda_i >0
\label{eq:constraints}
\end{align}
Here, we rewrite the strain tensor $u^{(1)}_j$ as the gradient of the displacement field with proper coefficients represented by $\mathcal{P}_{jk \ell}$. 
%$\mathcal{O}_{ij}\mathbf{u}^{(1)}_j =0$.
If a displacement field satisfies all the constraints [Eq~\eqref{eq:constraints}], it is a zero mode. 
For a rigid solid with positive-definite $\mathcal{K}^{(1,1)}$, all the eigenvalues are positive and thus we have $d_c=d(d+1)/2$ constraints. On the other hand, the number of degrees
of freedom is $d$, because the displacement field has $d$ components. Thus, above 1D, a rigid body is always over-constraint ($d_c>d$). This conclusion 
is in agreement with the generalized counting argument for continuous media~\cite{sun2019maxwell} as well as the embedding theorem of Janet and Cartan~\cite{janet1926possibilite,cartan1927possibilite}.

As shown in the main text, for systems at the verge of mechanical stability, some of the $\lambda_i$'s may vanish. If $d(d-1)/2$ of them vanish and the remaining $d$ eigenvalues
are positive, the system is at the Maxwell point, with the same number of constraints and degrees of freedom, within this leading-order theory.

\subsection{Second order terms}
The second-order derivatives of the displacement field form a rank-3 tensor 
$\mathbf{u}^{(2)}=\{ \partial_i \partial_j u_k\}$, and the first two indices in this tensor are symmetric  because $\partial_i \partial_j u_k=\partial_j \partial_i  u_k$.
As a result, $\mathbf{u}^{(2)}$ has $d^2 (d+1)/2=d d_c$ components.

By considering both first- and second-order derivative terms, the elastic energy now takes the following form
\begin{align}
E=\frac{1}{2} \int \mathbf{dr} \; (\mathcal{K}_{ij}^{(1,1)} u_i^{(1)} u_j^{(1)}
+2\mathcal{K}_{ij}^{(1,2)} u_i^{(1)} u_j^{(2)}
+\mathcal{K}_{ij}^{(2,2)} u_i^{(2)} u_j^{(2)})
\end{align}
where we used Einstein's notation to sum over repeated indices and the middle term is the cross term between $\mathbf{u}^{(1)}$ and $\mathbf{u}^{(2)}$.
As shown in the previous section, $\mathcal{K}^{(1,1)}$ is a $d_c\times d_c$ matrix. $\mathcal{K}^{(2,2)}$ and $\mathcal{K}^{(1,2)}$
are $(d d_c)\times (d d_c)$ and $d_c\times (d d_c)$ matrices respectively.

Same as shown in the previous section, the $\mathcal{K}^{(1,1)}$ matrix can be diagonalized by an orthogonal matrix $\mathcal{O}$.
With the help of this orthogonal matrix, the elastic energy can be written as
\begin{align}
E=\frac{1}{2} \int \mathbf{dr} \;  \left[\sum_{i=1}^{d_c}\lambda_{i}  (\mathcal{O}_{ij}u^{(1)}_j)^2
+2 (\mathcal{K}_{kl}^{(1,2)}\mathcal{O}^T_{ki}  u_l^{(2)})  (\mathcal{O}_{ij}u^{(1)}_j)
+\mathcal{K}_{ij}^{(2)} u_i^{(2)} u_j^{(2)}\right]
\end{align}
In the second term, we used the fact that $\mathcal{O}$ is an orthogonal matrix and thus $\mathcal{O}^T_{ki}\mathcal{O}_{ij}=\delta_{kj}$.

As demonstrated  in Sec.~\ref{app:sec:simple:exmpale}, for $\lambda_i>0$, we can absorb $2 (\mathcal{K}_{kl}^{(1,2)}\mathcal{O}^T_{ki}  u_l^{(2)})  (\mathcal{O}_{ij}u^{(1)}_j)$
into $\lambda_{i}  (\mathcal{O}_{ij}u^{(1)}_j)^2$, i.e.,
\begin{align}
\lambda_{i}  (\mathcal{O}_{ij}u^{(1)}_j)^2+2 (\mathcal{K}_{kl}^{(1,2)}\mathcal{O}^T_{ki}  u_l^{(2)})  (\mathcal{O}_{ij}u^{(1)}_j)
=\lambda_{i}  \left(\mathcal{O}_{ij}u^{(1)}_j+ \frac{\mathcal{K}_{kl}^{(1,2)}\mathcal{O}^T_{ki}}{\lambda_i}  u_l^{(2)}\right)^2
-\frac{\mathcal{K}_{mn}^{(1,2)}\mathcal{O}^T_{mi}\mathcal{K}_{kl}^{(1,2)}\mathcal{O}^T_{ki}}{\lambda_i}u_l^{(2)}u_n^{(2)}.
\end{align}
Here, we don't sum over $\lambda_i$, but all other repeated indices are summed over.
For $\lambda_i=0$, the corresponding cross term $2 (\mathcal{K}_{kl}^{(1,2)}\mathcal{O}^T_{ki}  u_l^{(2)})  (\mathcal{O}_{ij}u^{(1)}_j)$ must vanish, because this quadratic form is 
semi-positive definite. In summary, we can always rewrite the elastic energy as
\begin{align}
E=\frac{1}{2} \int \mathbf{dr} \;  \left[\sum_{\lambda_i>0}\lambda_{i}  (\mathcal{O}_{ij}u^{(1)}_j+ \frac{\mathcal{K}_{kl}^{(1,2)}\mathcal{O}^T_{ki}}{\lambda_i}  u_l^{(2)})^2
+(\mathcal{K}_{ln}^{(2)}
-\sum_{\lambda_i>0}\frac{\mathcal{K}_{mn}^{(1,2)}\mathcal{O}^T_{mi}\mathcal{K}_{kl}^{(1,2)}\mathcal{O}^T_{ki}}{\lambda_i})u_l^{(2)}u_n^{(2)}
\right] .
\end{align}
Up to the second-order derivative terms, this is precisely the structure in Eq.~\eqref{app:eq:elastic:full}.
We can repeat the same procedure as higher order derivative terms are included into the elastic energy, which leads to  Eq.~\eqref{app:eq:elastic:full} and ~\eqref{app:eq:elastic:energy:block:matrix}.

\subsection{States of self stress and the Maxwell-Calladine index theorem}
In this section, we show that similar to discrete systems, the transpose of the $\mathcal{C}$ matrix in the continuum theory dictates states of self stress.

As shown in the main text, the elastic energy in the continuum theory  is 
\begin{align}
E=\frac{1}{2}\int \mathbf{dr}\; \mathbf{u}^T\overleftarrow{\mathcal{C}}^T \Lambda \overrightarrow{\mathcal{C}}\mathbf{u} ,
\label{app:eq:elastic:energy:c:matrix}
\end{align}
and the null space of $\mathcal{C}$ (i.e.,  every $\mathbf{u}$ that satisfies $\mathcal{C} \mathbf{u}=0$) gives zero modes. In the main text, only the first set of squares 
in Eq.~\eqref{app:eq:elastic:full} (with coefficients $\lambda_i$) was considered. 
But Eq.~\eqref{app:eq:elastic:energy:c:matrix} can be easily generalized to  the full elastic energy (i.e. the entire Eq.~\eqref{app:eq:elastic:full} including all terms with coefficients
$\lambda_i$, $\widetilde{\lambda}_i$, $\widetilde{\widetilde{\lambda}}_i$, $\ldots$). This can be achieved by adding more rows to the $\mathcal{C}$ matrix, and each row of the 
$\mathcal{C}$ matrix corresponds to one square (with a nonzero coefficient) in Eq.~\eqref{app:eq:elastic:full}, which enforces one constraint to the system. 
If all the constraints are satisfied, $\mathcal{C} \mathbf{u}=0$, the displacement field is a zero mode.

Here, we study the  null space of $\mathcal{C}^T$, which is also known as the left null space of $\mathcal{C}$, and show that it gives 
the states of self stress. Assume that there exists a $\mathbf{w}$ such that
\begin{align}
\mathcal{C}^T \mathbf{w}=0 .
\end{align}
With this $\mathbf{w}$, the system has a state of self stress. Under this self stress, the elastic energy is
\begin{align}
E=\frac{1}{2}\int \mathbf{dr}\; 
\left(\Lambda^{-1}\mathbf{w}+\mathcal{C}\mathbf{u}\right)^T
\Lambda 
\left(\Lambda^{-1}\mathbf{w}+\mathcal{C}\mathbf{u}\right) .
\end{align}
%where $\Lambda^{-1}$ is the inverse of $\Lambda$. 
With $\mathcal{C}^T \mathbf{w}=0$, it is easy to check that the cross term vanishes (i.e., $\mathbf{u}^T\mathcal{C}^T\mathbf{w}=0$). Thus, up to a constant 
$\int \mathbf{dr} \; \mathbf{w}^T\Lambda^{-1}\mathbf{w}/2$, this elastic energy is identical to Eq.~\eqref{app:eq:elastic:energy:c:matrix}.

It is easy to check that this elastic energy describes a system under stress, i.e., the stress
\begin{align}
\sigma^{(m)}=\frac{\delta E}{\delta \mathbf{u}^{(m)}} \ne 0 
\end{align}
%e.g. the stress tensor
%\begin{align}
%\sigma_{ij}=\frac{\delta E}{\delta u_{ij}} \ne 0 
%\end{align}
even at zero deformation. %, where $u_{ij}=(\partial_i u_j+\partial_j u_i)/2$ is the strain tensor.
Here,  because the elastic energy contains higher-order gradients of the displacement field, their corresponding high-order stress fields are introduced here $\sigma^{(m)}$.
For $m=1$, $\sigma^{(m)}$ recovers the standard stress tensor. It is easy to verify that based on cutting stability (as defined above), a stress-free system must have
$\sigma^{(m)}=0$ for all $m$. And thus, $\mathbf{w}$ indeed induces stress in an elastic system. More importantly, this system is under perfect force balance, i.e., the force
\begin{align}
\mathbf{f}_{\mathbf{u}=0}=-\frac{\delta E}{\delta \mathbf{u}}\bigg |_{\mathbf{u}=0}= - \mathcal{C}^T\mathbf{w}=0
\end{align}
Thus this is a state of self stress. Because every $\mathbf{w}$ defines a state of self stress. 
the number of states of self stress is determined by the dimension of the null space of $\mathcal{C}^T$.

Because the null spaces of $\mathcal{C}$ and $\mathcal{C}^T$ give zero modes and states of self-stress, the Maxwell-Calladine theorem still holds in the continuum.
This theorem has been proved in discrete lattices/networks. Our continuum theorem now generalizes it to continuous media. Here we briefly 
outline the proof in the continuum.
As mentioned in the main text and above, the number of rows of $\mathcal{C}$ is the number of constraints $N_c$, while the number of columns is the number of degrees of freedom $N_d$.
Thus, $\mathcal{C}$ is a $N_c\times N_d$ matrix, while its transpose $\mathcal{C}^T$ is  $N_d\times N_c$.
For an arbitrary matrix, the dimension of the null space is the the number of columns minus the rank of the matrix. Thus, the  null space of $\mathcal{C}$ has dimension
$N_d- \rank{(\mathcal{C})}$, while the null space of $\mathcal{C}^T$ has $N_c- \rank{(\mathcal{C}^T)}$. 
Because the dimension of the null space of $\mathcal{C}$ ($\mathcal{C}^T$) is the number of zero modes (states of self stress), this means that
\begin{align}
N_0= &N_d- \rank{(\mathcal{C})} ,
\\
N_s=& N_c- \rank{(\mathcal{C}^T)} ,
\end{align}
where $N_0$ and $N_s$ represent the numbers of zero modes and states of self stress respectively.
Because the rank of a matrix always equals to the rank of its transpose, $ \rank{(\mathcal{C})}=\rank{(\mathcal{C}^T)}$, the two equations above imply
that
\begin{align}
N_0-N_s= N_d-N_c ,
\end{align}
which is known as the Maxwell-Calladine index theorem. In sharp contrast to lattice systems, here $N_d$, $N_s$, $N_0$ and $N_s$ are numbers in the continuum theory and 
are independent of microscopic details.
For example, in the continuum theory constructed above (based on displacement field gradients), $N_d$ is the number of spatial dimensions, regardless of lattice structures and/or 
how many nodes a lattice unit cell have. While in  lattice systems, its number of degrees of freedom per unit cell is $d \times N_n$, where $N_n$ is the number of nodes in a cell. 
The lattice degrees of freedom is very sensitive to lattice structure, and its number can become very big, if each unit cell contains a high number of nodes 
(and the number of nodes per cell diverges in a continuum medium), while in contrast, the continuum theory $N_d$ is always finite ($d$) and its value is independent of microscopic details.

\subsection{Low-energy optical modes}
In this section, we consider an elastic system that contains soft optical phonon modes, i.e., optical phonon modes with a small gap. 
If the gap of such a optical phonon mode (at zero wavevector $q=0$) is zero, (e.g., in a regular kagome lattice with nearest springs~\cite{Sun2012}),
the energy of this optical mode is comparable to (and even lower than) the acoustic ones, and thus it must be included in the continuum theory to correctly describe 
the low-energy physics. Generically, such a soft optical phonon mode has a small but finite gap (e.g., twisted or deformed kagome lattices~\cite{Sun2012, Kane2014}). 
For these gapped soft optical modes, their contributions to low-energy physics becomes negligible in the long-wave-length (small $q$) and low-energy (small $\omega$) limit. 
And thus if we focus on small wavevector ($q$ close to $0$), the continuum theory do not need to include these optical modes.
However, at intermediate wavevector, at which the energy of the soft optical mode becomes comparable with that of the acoustic ones, 
including these soft modes provides a more accurate description. 

If we include $n_O$ such soft optical modes in the continuum theory, $\phi_i$ with $i=1$, $2$, $\ldots$, $n_O$, we just need to 
introduce some more components for each $\mathbf{u}^{(m)}$. 
%these modes $\phi_i$ (without derivatives) to $\mathbf{u}^{(1)}$ as its components, and the first order derivatives  $\partial_i \phi_j$ to $\mathbf{u}^{(2}$, etc.
%In other words, 
For example, $\mathbf{u}^{(1)}$ is now a $[d(d+1)/2+n_O]$-component vector. The first $d(d+1)/2$ components are the strain tensor (same as before), and in addition
it now contains $n_O$ more components, which are these soft optical modes $\phi_i$. We do the same for all $\mathbf{u}^{(m)}$, i.e., introducting the $(m-1)$th order
derivatives of $\phi_i$ as new components of $\mathbf{u}^{(m)}$. Then, we can follow the same procedure shown above to reorganize the
elastic energy and to define the $\mathcal{C}$ matrix.

\section{Low-energy theory}
In this section, we present a generic framework to compute coefficients in  Eq.~\eqref{app:eq:elastic:full} from first principles. 
This objective is achieved through integrating out high energy degrees of freedom (i.e., optical modes). In a continuum theory of elasticity,
this is a standard procedure. However,  the conventional approach usually focuses on the dynamic matrix~\cite{Mao2011a}, which doesn't carry topological information 
about Maxwell systems. Here, we develop a more general approach, which can be applied to both the $\mathcal{C}$ matrix and
the dynamic matrix. For the dynamic matrix, it gives the same results as the conventional approach. 
Remarkably, as far as zero modes are concerned, although our method dramatically reduces the number of degrees of freedom,
it contains the same zero modes identical to the full theory.

\subsection{Elastic theory in the real space}
In this section we consider an elastic medium with a periodic structure. It must be emphasized that here we don't assume the system is an ideal lattice model composed of ideal 
springs. Instead, the scope of the study also includes more realistic setups used in experiments, which requires finite element analysis to describe.

In an ideal discrete lattice model, the displacement field is defined on a set of discrete nodes.
For a more-realistic continuous medium, we can still utilize a discrete set of nodes to describe the system through the finite-element method (FEM).
For elastic properties, the FEM is a reliable approximation, which becomes asymptotically exact as the mesh becomes denser and denser.
Below, we will use this discrete setup to define and compute the topological index. 
Because a topological index measures the topological property of a system, which is a qualitative property insensitive to quantitative details,
as long as the mesh is not too sparse, the FEM is expected to produce the exact value of the topological index.

Here, we consider an infinite-largely periodic structure.  In each unit cell, we introduce the same set of discrete nodes, which will be labeled by an integer
$\alpha=1$, $2$, $\ldots$, $N_n$ with $N_n$ being the number of nodes per cell. We will label each unit cell by its real-space coordinate $\mathbf{R}$, 
which is a lattice vector  $\mathbf{R}=\sum_{i=1}^{d} n_i \mathbf{a}_i$ with $n_i$ being integers and $\{\mathbf{a}_i\}$ representing a set of primitive lattice vectors.
With this setup, the deformation of a node is represented by
\begin{align}
\mathbf{u}^{(\alpha)}(\mathbf{R})= \left(u_1^{(\alpha)}(\mathbf{R}),u_2^{(\alpha)}(\mathbf{R}),\ldots,u_d^{(\alpha)}(\mathbf{R})\right) .
\end{align}
Here, $\mathbf{u}$ is the displacement vector of the $\alpha$th node in the unit cell located at $\mathbf{R}$. Because $\mathbf{u}$ is a $d$-dimensional vector, below we will use
a subindex $i=1$, $\ldots$, $d$ to label its components.

Without loss of generality, here we assume the mass matrix is an identity matrix, which can always be achieved by changing the basis. As a result, 
the frequency of a mode is just the square root of the energy.
With a periodic structure, the elastic energy of the system can be written as
\begin{align}
E=\sum_{\mathbf{R}} E_{\mathbf{R}} ,
\label{app:eq:E:as:ER}
\end{align}
where $E_{\mathbf{R}}$ is the elastic energy associated with the unit cell located at $\mathbf{R}$.
Due to the lattice translational symmetry, the elastic energy associated with two different unit cells (e.g. $E_{\mathbf{R}_a}$ and $E_{\mathbf{R}_b}$) are related by a lattice transition with a translation vector $\mathbf{t}=\mathbf{R}_b-\mathbf{R}_a$.
It is also worthwhile to mention that $E_{\mathbf{R}}$ depends not only on nodes inside the unit cell $\mathbf{R}$, but also nodes from neighboring cells, because the elastic energy
in general contains cross terms between nodes in different unit cells.

Within linear elasticity, each $E_{\mathbf{R}}$ is a quadratic form of the displacement field, and for a stress-free system, it is semi-positive definite.
For a semi-positive-definite quadratic form, it can always be written as the sum of a set of squares. 
\begin{align}
E_\mathbf{R}=\frac{1}{2}    \sum_m \kappa_m \left[\xi_m(\mathbf{R})\right]^2 ,
\label{app:eq:square:form:ER}
\end{align}
where $\kappa_m>0$ are positive coefficients and $\xi_m$ are linear combinations of the displacement field
\begin{align}
\xi_m(\mathbf{R)}=\sum_{\mathbf{t},i} \mathcal{M}_{m i}^{(\alpha)}(\mathbf{t})u^{(\alpha)}_i(\mathbf{R}+\mathbf{t}) .
\label{app:eq:xi:real:space}
\end{align}
Here, $\mathcal{M}_{m i}^{(\alpha)}(\mathbf{t})$ are some real coefficients and $\mathbf{t}$ is a lattice vector. For nodes in the unit cell $\mathbf{R}$, $\mathbf{t}=0$, but 
because nodes
from neighboring cells will also arise here, it is necessary to include nonzero $\mathbf{t}$ in $\xi_m(\mathbf{R})$.

Mathematically, the reason that we can always written $E_\mathbf{R}$ in this square sum form [Eq.~\eqref{app:eq:square:form:ER}] is because each quadratic form can be 
represented by a real symmetric matrix, and once we diagonalize the matrix and use its eigenvectors as the new basis to rewrite the quadratic form, it turns into the sum 
of a set of squares.
Physically, from the view point of Maxwell's counting argument, each square term (with a positive $\kappa_m$)
enforces one constraint. In an ideal model composed of balls connected by ideal springs, 
each square term corresponds to the elastic energy of one spring and $\kappa_m$ is the spring constant. 
%For a more realistic model, the physical meaning  of each square term corresponds to one normal mode of elastic deformations.

\subsection{Wave vector space formula}\label{sec:kspace}
Now, we go to the wave vector space via a Fourier transformation, 
\begin{align}
\mathbf{u}^{(\alpha)}(\mathbf{q})&=\sum_{\mathbf{R}} \mathbf{u}^{(\alpha)} (\mathbf{R}) \exp(-i \mathbf{q}\cdot \mathbf{R}) ,
\label{app:eq:ft}
\\
\mathbf{u}^{(\alpha)}(\mathbf{R})&=\frac{1}{\Omega_{B.Z.}}\int_{B.Z.} {d\mathbf{q}} \; \mathbf{u}^{(\alpha)} (\mathbf{q}) \exp(i \mathbf{q}\cdot \mathbf{R}) .
\label{app:eq:ft:inverse}
\end{align}
Here $\mathbf{u}^{(\alpha)} (\mathbf{q})$ is the displacement of the node $\alpha$ in the wave vector space, while $\mathbf{u}^{(\alpha)} (\mathbf{R})$ is the real space displacement of the node $\alpha$ in the unit cell $\mathbf{R}$. 
In Eq.~\eqref{app:eq:ft}, we sum over all lattice vectors, and the integral in Eq.~\eqref{app:eq:ft:inverse} is over one Brillouin zone with $\Omega_{B.Z.}$ beging 
the volume of a Brillouin zone. 

Notice that in this Fourier transformation, the phase factor is chosen to be $\exp(-i \mathbf{q}\cdot \mathbf{R})$ with $\mathbf{R}$ being a lattice vector.
This convention has two consequences. First of all, it ensures that all functions defined in the wave vector space are periodic with respect to the Brillouin zone, e.g., 
$\mathbf{u}^{(\alpha)}(\mathbf{q}+\mathbf{G})=\mathbf{u}^{(\alpha)}(\mathbf{q})$ where $\mathbf{G}$ is a reciprocal lattice vector. 
This periodicity in the wave vector space is crucial for the definition of the topological index, same as in lattice systems.
Secondly, it is easy to notice that this phase factor of the Fourier transformation depends on how we choose the unit cell. 
For a periodic structure, there exist infinitly many equivalent ways to define a unit cell~\cite{kittel1976introduction}. 
Each option is known as a gauge choice. 
Such a  gauge choice will not change any physical observables, i.e., physics observable should be gauge invariant.
However, if we choose a different unit cell, the same node may now be assigned to a different cell (with a different $\mathbf{R}$) and thus the corresponding phase factor 
$\exp(-i \mathbf{q}\cdot \mathbf{R})$ will be shifted. This phase factor turns out to play a very important role in the definition of the topological index, and most importantly, 
for Maxwell lattices, the conventional way of computing the topological index is sensitive to the gauge choice, and different gauge choices will shift the index by an integer constant.
As a result, if we want to connect the topological index with a physics observable, e.g., number of edge modes, it is necessary to deduct the gauge sensitive part. 
%Below, we introduce a gauge-independent definition of the topological index, which works for both lattice systems and in the continuous theory.

In the wave vector space, the elastic energy becomes
\begin{align}
E=\frac{1}{\Omega_{B.Z.}}\int_{B.Z.} {d\mathbf{q}} \; E (\mathbf{q}) ,
\end{align}
where 
\begin{align}
E(\mathbf{q})=\frac{1}{2}\sum_m \kappa_m |\xi_m(\mathbf{q})|^2 .
\end{align}
The coefficients $\kappa_m$ are defined in Eq.~\eqref{app:eq:square:form:ER}, and $\xi_m(\mathbf{q})$
is the wave vector space version of Eq.~\eqref{app:eq:xi:real:space}
\begin{align}
\xi_m(\mathbf{q})=\sum_{\mathbf{t},i} \mathcal{M}_{m i}^{(\alpha)}(\mathbf{t}) \; \exp(-i \mathbf{q}\cdot\mathbf{t}) \; u^{(\alpha)}_i(\mathbf{q}).
\end{align}
This elastic energy can be written in a matrix form. Here, we write all displacement field (in the wave vector space) as a column vector 
$\mathbf{u}(\mathbf{q})=\{ u^{(\alpha)}_i(\mathbf{q}) \}$, and 
define
%a diagonal matrix $\mathcal{K}$ using the coefficients $ \kappa_m$. In addition, we introduce 
a matrix $\mathcal{C}(\mathbf{q})$ as
\begin{align}
\mathcal{C}(\mathbf{q})_{m,(\alpha)i}=\sqrt{\kappa_m} \sum_{\mathbf{t}} \mathcal{M}_{m i}^{(\alpha)}(\mathbf{t}) \; \exp(-i \mathbf{q}\cdot\mathbf{t})
\label{app:eq:cmatrix:kspace}
\end{align}
where the row index $m$ runs over all the square terms [defined in Eq.~\eqref{app:eq:square:form:ER}], and the column index $(\alpha)i$ goes over
all the components of the displacement vectors for all nodes. Notice that here we absorbed $\kappa_m$ into the definition of  $\mathcal{C}(\mathbf{q})$, which
differs slightly from the convention used in Maxwell lattices. There, a separate matrix is utilized for $\kappa_m$, which corresponds to the spring constant and is 
usually set to unity.
%$\sum_{\mathbf{t},i} \mathcal{M}_{m i}^{(a)}(\mathbf{t}) \; \exp(i \mathbf{q}\cdot\mathbf{t})$. 

With the $\mathcal{C}$ matrix defined above, the elastic energy can be written as
\begin{align}
E(\mathbf{q})=\frac{1}{2} \mathbf{u}(\mathbf{q})^\dagger \mathcal{C}(\mathbf{q})^\dagger  \mathcal{C}(\mathbf{q} )\mathbf{u}(\mathbf{q}) .
\label{app:eq:Ek}
\end{align}
From this formula, it is easy to realize that the dynamic matrix $\mathcal{D}({\mathbf{q}})$ is 
\begin{align}
\mathcal{D}(\mathbf{q})= \mathcal{C}(\mathbf{q})^\dagger  \mathcal{C}(\mathbf{q}).
\label{app:eq:dmatrx:cmatrix}
\end{align}
For an idea lattice, this $\mathcal{C}(\mathbf{q})$ recovers the $\mathcal{C}$ matrix defined for discrete lattices, and the matrix becomes a square matrix 
for a Maxwell lattice. For a continuous medium, in general $\mathcal{C}(\mathbf{q})$ is not a square matrix. Instead, we expect the matrix to have more
rows than columns, which implies that the system is over-constrained at the microscopic level. 

It is worthwhile to emphasize here that same as in ideal lattice models, the $\mathcal{C}(\mathbf{q})$ matrix contains more information than the dynamic matrix. 
As shown in Eq.~\eqref{app:eq:dmatrx:cmatrix},  $\mathcal{C}(\mathbf{q})$  uniquely determines the dynamic matrix. However, from the dynamic matrix, although it can be mathematically decomposed into the form of Eq.~\eqref{app:eq:dmatrx:cmatrix}, the decomposition
is not unique and thus  $ \mathcal{C}(\mathbf{q})$ cannot be uniquely identified from $\mathcal{D}({\mathbf{q}})$.
Below, we will keep working with 
the $\mathcal{C}$ matrix, without condense its information into $\mathcal{D}$, which is crucial for the topological index.

\subsection{Separating high- and low- energy modes.}
For simplicity, we now drop the wave vector $(\mathbf{q})$ used in Eqs.~\eqref{app:eq:Ek} and~\eqref{app:eq:dmatrx:cmatrix}.
Here, we split displacement fields $\mathbf{u}({\mathbf{q}})$ into the low-energy and high-energy parts. 
This is achieved via changing of basis. We diagonalize the dynamical matrix at $\mathbf{q}=0$ and using its eigenvectors as the new basis. 
We emphasize here that the eigenvectors utilized here are obtained at $\mathbf{q}=0$, and thus this changing of basis is $q$-independent.
Here, we choose to order
the eigenvectors according to their eigenvalues from low to high. The first $d$ eigenvectors here are the acoustic modes with eigenvalues $0$. The rest are
optical modes, which are usually gapped.  Because the continuum theory focuses on small wave vectors ($\mathbf{q}$ close to $0$), this new basis
naturally separates low- and high-energy modes
\begin{align}
\mathbf{u}=\begin{pmatrix}
\mathbf{u}_L\\
\mathbf{u}_H
\end{pmatrix}
\label{app:eq:mode:split}
\end{align}
Here, the first a few components will be our low-energy modes. Typically, we keep the lowest $d$ modes, i.e., acoustic modes. However, in some systems, especially systems
near the verge of mechanical stability, the energy of certain optical modes may be close to zero. These soft optical modes can also be included in $\mathbf{u}_L$ as well.
As shown in the main text, including soft optical modes can help improve the accuracy of the continuum theory, especially near the wave vector at which 
the energy of the soft optical mode becomes comparable with acoustic modes.

Once we split the displacement fields into the low- and high-energy sectors, the $\mathcal{C}$ matrix define in Eq.~\eqref{app:eq:cmatrix:kspace} is also splitted into two blocks
\begin{align}
\mathcal{C}=
\begin{pmatrix}
\mathcal{A}& \mathcal{B}
\end{pmatrix} .
\label{app:eq:C:split}
\end{align}
The matrices $\mathcal{A}$ and $\mathcal{B}$ have the same number of rows as  $\mathcal{C}$ (i.e. same number of constraints), while the number of columns of $\mathcal{A}$ 
($\mathcal{B}$) is the dimensions of $\mathbf{u}_L$ ($\mathbf{u}_H$). With this block form, the elastic energy [Eq.~\eqref{app:eq:Ek}] 
becomes
\begin{align}
E(\mathbf{q})=&\frac{1}{2} 
\begin{pmatrix}
\mathbf{u}_L^\dagger & \mathbf{u}_H^\dagger
\end{pmatrix}
\begin{pmatrix}
\mathcal{A}^\dagger
\\
\mathcal{B}^\dagger
\end{pmatrix}
\begin{pmatrix}
\mathcal{A}& \mathcal{B}
\end{pmatrix}
\begin{pmatrix}
\mathbf{u}_L \\ \mathbf{u}_H
\end{pmatrix}
%%%%%%
=\frac{1}{2} 
\begin{pmatrix}
\mathbf{u}_L^\dagger & \mathbf{u}_H^\dagger
\end{pmatrix}
\begin{pmatrix}
\mathcal{A}^\dagger \mathcal{A} & \mathcal{A}^\dagger \mathcal{B}
\\
 \mathcal{B}^\dagger \mathcal{A} & \mathcal{B}^\dagger \mathcal{B}
\end{pmatrix}
\begin{pmatrix}
\mathbf{u}_L \\ \mathbf{u}_H
\end{pmatrix}
\nonumber\\
%%%%%%%%
=&\frac{1}{2} \big(\mathbf{u}_L^\dagger \mathcal{A}^\dagger \mathcal{A}\mathbf{u}_L
+\mathbf{u}_L^\dagger \mathcal{A}^\dagger \mathcal{B}\mathbf{u}_H
+\mathbf{u}_H^\dagger \mathcal{B}^\dagger \mathcal{A}\mathbf{u}_L
+\mathbf{u}_H^\dagger \mathcal{B}^\dagger \mathcal{B}\mathbf{u}_H
\big) .
\label{app:eq:block:Ek}
\end{align}

In Eq.~\eqref{app:eq:block:Ek}, the low- and high- energy modes are mixed together through the cross terms in the elastic energy. 
These cross terms can be absorbed into the high energy mode through defining a new basis, and the elastic energy [Eq.~\eqref{app:eq:block:Ek}] becomes
\begin{align}
E(\mathbf{q})=\frac{1}{2} \mathbf{u}_L^\dagger \mathcal{\widetilde{A}}^\dagger\mathcal{\widetilde{A}} \mathbf{u}_L+
\frac{1}{2}
\widetilde{\mathbf{u}}_H^\dagger\mathcal{B}^\dagger\mathcal{B} \widetilde{\mathbf{u}}_H ,
\label{app:eq:split:energy}
\end{align}
where
\begin{align}
\mathcal{\widetilde{A}}=
\mathcal{A}- \mathcal{B} (\mathcal{B}^\dagger \mathcal{B})^{-1} \mathcal{B}^\dagger \mathcal{A} ,
\end{align}
and
\begin{align}
\widetilde{\mathbf{u}}_H=\mathbf{u}_H+  (\mathcal{B}^\dagger \mathcal{B})^{-1}  \mathcal{B}^\dagger \mathcal{A} \mathbf{u}_L .
\end{align}
Here, we used the fact $(\mathcal{B}^\dagger \mathcal{B})$ is invertible. As shown in Eq~\eqref{app:eq:block:Ek}, in the dynamical matrix,
$(\mathcal{B}^\dagger \mathcal{B})$ is part of the dynamic matrix associated with the high-energy modes. Because these modes are gapped
optical phonons, $(\mathcal{B}^\dagger \mathcal{B})$ must be positive definite and thus it is invertible.

This new basis decouples the low- and high-energy modes. Thus, for low-energy physics, we can now focus on the low-energy sector and set
\begin{align}
E(\mathbf{q}) \simeq \frac{1}{2} \mathbf{u}_L^\dagger \mathcal{\widetilde{A}}^\dagger\mathcal{\widetilde{A}} \mathbf{u}_L .
\label{app:eq:reduced:E}
\end{align}
In comparison with Eq.~\eqref{app:eq:Ek}, here the $\mathcal{C}$ matrix is reduced into its low-energy sector $\widetilde{\mathcal{A}}$, while the degrees of freedom
is reduced from all displacement fields $\mathbf{u}$ into low-energy deformations only $\mathbf{u}_L$.
The physical meaning of this reduction in degrees of freedom can be viewed as integrating out high-energy modes, which is a standard procedure in elastic theory, especially
in the continuum theory. However, the conventional approach usually reduces the dynamic matrix into its low-energy sector. Here, we generalized
the approach to the $\mathcal{C}$ matrix, making it possible to study topological properties via low-energy degrees of freedom. 

Another way to understand this reduction is by trying to find low-energy deformations via minimizing the elastic energy. From Eq~\eqref{app:eq:split:energy}, it
is easy to notice that this minimization can be achieved by setting $\widetilde{\mathbf{u}}_H$ to zero, and then search for low-energy excitations in the reduced theory
[Eq.~\eqref{app:eq:reduced:E}].

To conclude this section, we emphasize  that this $\mathcal{\widetilde{A}}$ is \emph{not} a square matrix, in direct contrast to the $\mathcal{C}$ matrix utilized in discrete 
Maxwell lattices. This is because our reduction of degrees of freedom only removes high-energy degrees of freedom, while all the constraints are maintained
in the reduced theory. As a result,  $\mathcal{\widetilde{A}}$ and $\mathcal{C}$ matrices have the same number of rows (i.e., the same number of constraints), but  
$\mathcal{\widetilde{A}}$  contains fewer number of columns (i.e., fewer degrees of freedom).

\subsection{Exactly solution for zero modes}
In this section, we show that there exists a one-to-one correspondence 
between zero modes in the reduced theory, i.e., $\mathcal{\widetilde{A}} \mathbf{u}_L=0$ and zero modes in the full theory $\mathcal{C} \mathbf{u}=0$.
Therefore, as far as zero modes are concerned, our reduced theory offers the exact solutions with no error or approximation. 

If $\mathbf{u}=(\mathbf{u}_L, \mathbf{u}_H)$ is a zero mode satisfying $\mathcal{C} \mathbf{u}=0$, we have
\begin{align}
\mathcal{C} \mathbf{u}=\mathcal{A} \mathbf{u}_L+\mathcal{B} \mathbf{u}_H=0 .
\label{app:eq:zero:condition}
\end{align}
By multiplying to both sides $ (\mathcal{B}^\dagger \mathcal{B})^{-1}  \mathcal{B}^\dagger$, we get
$\mathbf{u}_H=-(\mathcal{B}^\dagger \mathcal{B})^{-1}  \mathcal{B}^\dagger \mathcal{A} \mathbf{u}_L$. 
If we substituted this $\mathbf{u}_H$ into Eq.~\eqref{app:eq:zero:condition}, we get
\begin{align}
[\mathcal{A}-(\mathcal{B}^\dagger \mathcal{B})^{-1}  \mathcal{B}^\dagger \mathcal{A}]  \mathbf{u}_L= \mathcal{\widetilde{A}} \mathbf{u}_L=0 .
\end{align}
Thus, $\mathcal{C} \mathbf{u}=0$ must implies $\mathcal{\widetilde{A}} \mathbf{u}_L=0$.

We now start from a zero mode in the reduced theory with $\mathcal{\widetilde{A}} \mathbf{u}_L=0$.
By setting $\mathbf{u}_H=-(\mathcal{B}^\dagger \mathcal{B})^{-1}  \mathcal{B}^\dagger \mathcal{A} \mathbf{u}_L$, it is easy to check that
\begin{align}
\mathcal{C} \mathbf{u}=\mathcal{A} \mathbf{u}_L+\mathcal{B} \mathbf{u}_H=
\mathcal{A} \mathbf{u}_L+\mathcal{B}
[\mathcal{A}-(\mathcal{B}^\dagger \mathcal{B})^{-1}  \mathcal{B}^\dagger \mathcal{A}] \mathbf{u}_L
=\mathcal{\widetilde{A}} \mathbf{u}_L
=0
\end{align}

\subsection{The continuum theory}
As mentioned in the previous section, the $\mathcal{\widetilde{A}}$ matrix is not yet the $\mathcal{C}$ matrix of the continuum theory. One more procedure is still
needed to get the coefficients in Eq.~\eqref{app:eq:elastic:full}.

Here, we take a Taylor expansion and write $\mathcal{\widetilde{A}}$ as a power series of $\mathbf{q}$. For the continuum theory, although keeping more terms
improves its accuracy at larger $\mathbf{q}$,  two leading-order terms are usually sufficient as far as the topology at long-wavelength is concerned. 
For acoustic modes, the leading order theory is linear in $\mathbf{q}$, and we also need to keep the next order terms ($\propto \mathbf{q}^2$) as well. 
For soft-optical modes (if they are included in the low-energy sector), their leading order terms are constants independent of $\mathbf{q}$, and 
the next order terms, which are proportional to $\mathbf{q}$, are also needed.

To obtain the continuum theory, now we go back to the real space (replacing $\mathbf{q}$ with $-i \nabla$). Here, $\mathcal{\widetilde{A}} \mathbf{u}_L$ becomes
a linear combination of displacement fields and their derivatives. And in real space, the elastic energy is
\begin{align}
E= \frac{1}{2} \mathbf{u}_L^{T}  
\overleftarrow{\mathcal{\widetilde{A}}}^{T}\overrightarrow{\mathcal{\widetilde{A}}} \mathbf{u}_L
\label{app:eq:reduced:E:real:space}
\end{align}
where the arrow above $\mathcal{\widetilde{A}}$ implies that the derivatives in $\mathcal{\widetilde{A}}$ are for $\mathbf{u}_L$ on the right side of it, while the derivatives in
the transpose matrix $\mathcal{\widetilde{A}}^T$ are for $\mathbf{u}_L^{T}$  on the left side.

Now, Eq.~\eqref{app:eq:reduced:E:real:space} recovers the generic form of the elastic energy [Eq.~\eqref{app:eq:elastic:energy:block:matrix}], and we can follow the procedure
discussed in that section to reorganize the elastic energy into the form shown in Eq.~\eqref{app:eq:elastic:full}. 
As shown in the main text, from there, the $\mathcal{C}$ matrix and topological indices in the continuum theory can be defined.

\section{Topological index, zero modes and gauge choice}
In this section we discuss the connection between the topological index and zero modes, as well as how to make the index gauge independent.

Discussions in Sec.~\ref{app:sub:sec:zeros} applies to both Maxwell lattices and Maxwell media, while Secs.~\ref{app:sub:sec:lattices}, \ref{app:sub:sec:lattices:index}
and~\ref{app:sub:sec:lattices:dangling} focus on discrete systems. And Sec.~\ref{app:sub:sec:continuum} is devoted to the continuum theory.

\subsection{Real-space zero modes and zeros in $\det\mathcal{C}(\mathbf{q})$}
\label{app:sub:sec:zeros}
In this section, we use a discrete setup to demonstrate the physics, but the same conclusions can be easily generalized to the continuum theory
using the continuum elastic energy and its $\mathcal{C}$ matrix defined in the main text.
Here, we show that zero modes in the real space is related with zeros in the determinant of 
the wave vector space $\mathcal{C}$ matrix, $\det\mathcal{C}(\mathbf{q})=0$.

As shown above, for an elastic medium with a periodic structure, the total elastic energy can be written as the sum of $E_\mathbf{R}$, which 
is the elastic energy associated with each unit cell as shown in Eq.~\eqref{app:eq:square:form:ER}. 
%This is true for infinite systems and for systems without a boundary (e.g., under  periodic boundary conditions). In addition, this is also true for systems with open edges, as long as the edge doesn't cut through a unit cell (i.e. for unit cells at the boundary, they also have energy $E_\mathbf{R}$, instead of part of  $E_\mathbf{R})$. This can usually be achieved, if proper unit cell is chosen.

Here, we consider a deformation pattern, 
\begin{align}
\mathbf{u}=\mathbf{A} e^{i q_x x+ i q_y y}
\label{app:eq:wave:form}
\end{align}
where $q_x$ is real and $q_y$ is complex. This setup describes bulk waves (if $q_y$ is real)
and edge modes along edges parallel to the $x$ axis. From Eq.~\eqref{app:eq:square:form:ER}, the elastic energy for this deformation 
can be easily obtained
\begin{align}
 E_R =\frac{1}{2}\frac{ \mathbf{A}^\dagger [\mathcal{C}(\bar{q}_x, q'_y+i q''_y)]^\dagger \mathcal{C}(\bar{q}_x, q'_y+i q''_y)  \mathbf{A}}{\mathbf{A}^\dagger\mathbf{A}}
\end{align}
where $q_y'$ and $q_y''$ are the real and imaginary part of $q_y$, $q_y= q_y'+i q''_y$.
Although this formula may look like Eq.~\eqref{app:eq:Ek}, its physical meaning is different. Here, we are computing the elastic energy 
of a deformation in the real space, while Eq.~\eqref{app:eq:Ek} is a wave vector space formula. 
From this $\langle E_R \rangle$, it is easy  to realize 
that a zero energy mode must satisfy 
\begin{align}
\mathcal{C}(\bar{q}_x, q'_y+i q''_y)  \mathbf{A}=0.
\end{align}
For a Maxwell system, $\mathcal{C}(\bar{q}_x, q'_y+i q''_y)$ is a square matrix, and thus to have a nontrivial solution, we have to require
\begin{align}
\det\mathcal{C}(\bar{q}_x, q'_y+i q''_y) =0 .
\end{align}

\subsection{$\det\mathcal{C}$ in discrete lattices}
\label{app:sub:sec:lattices}
In this section, we focus on discrete Maxwell lattices and show that at a fixed $q_x$, as long as interactions are short-ranged, 
$\det\mathcal{C}(\bar{q}_x, q_y)$ must take the following functional form
\begin{align}
\det \mathcal{C} (z)=c\; z^N \;   \Pi_{i=1}^{M} (z-z_i)
\label{app:eq:poly:equation}
\end{align}
where $z=\exp(- i  2\pi  q_y/b_2)$ is a complex variable with $b_2$ being the length of a reciprocal lattice vector (see below for details). 
$c$ and $\{z_i \ne 0\}$ are complex constants, while $N$ and $M$ are integers. Here, $z_i$'s are nonzero and $M \ge 0$, while $N$ can be positive,
zero or negative.
As will shown below, this implies that as a function of $z$, the function utilized to define the topological index, $\tr \mathcal{C}^{-1} \partial_{z}\mathcal{C}$,
is an meromorphic function for the entire complex $z$ plane (i.e., analytic except for some discrete poles). In addition, the number of poles is finite, which 
equals to $M+2$. Among these $M+2$ poles, $M$ of them ($z=z_i$) correspond to zero modes and their contributions to the topological index are gauge invariant. 
The other $2$ poles ($z=0$ and $z=\infty$) don't corresponds to physical zero modes, and their contributions make the topological index gauge dependent.

Here, we use $\mathbf{a}_1$ and $\mathbf{a}_2$ to represent the two primitive lattice vectors, and  $\mathbf{b}_1$ and $\mathbf{b}_2$ as their corresponding 
reciprocal lattice vectors. For a Maxwell lattice, typically, the edge is chosen to be along a lattice vector, and here
we set the edge to be along  $\mathbf{a}_1$, which is our $x$ axis. Because the $y$ axis is perpendicular to the edge ($\mathbf{a}_1$), 
by definition, $\mathbf{b}_2$ must be along the $y$ axis.

The matrix elements of the $\mathcal{C}$ matrix are shown in Eq.~\eqref{app:eq:cmatrix:kspace}, and its $q_y$ dependence comes from the exponential 
$\exp(- i \mathbf{q}\cdot\mathbf{t})$, where $\mathbf{t}$ is a lattice vector $\mathbf{R}=m \mathbf{a}_1+n \mathbf{a}_2$. Thus
\begin{align}
\exp(- i \mathbf{q}\cdot\mathbf{t})=\exp\{- i [(m \mathbf{a}_1+ n \mathbf{a}_2)\cdot{\mathbf{\hat{x}}}]q_x\}
\exp\{- i [(m \mathbf{a}_1+ n \mathbf{a}_2)\cdot{\mathbf{\hat{y}}}]q_y\} .
\label{app:eq:phase:z:1}
\end{align}
Here we use the fact that $\mathbf{q}=q_x \mathbf{\hat x} +q_y  \mathbf{\hat y}$ with  $\mathbf{\hat x}$ and $\mathbf{\hat y}$ being unit vectors along $x$ and $y$.
We now focus on the second phase factor in Eq.~\eqref{app:eq:phase:z:1},
which is associated with $q_y$. Because $\mathbf{\hat y} \parallel \mathbf{b}_2$,  we have $\mathbf{\hat y}\cdot a_1=0$, and thus
\begin{align}
\exp\{- i [(m \mathbf{a}_1+ n \mathbf{a}_2)\cdot{\mathbf{\hat{y}}}]q_y\}=\exp(- i  2\pi n \frac{q_y}{b_2})=z^n .
\label{app:eq:phase:z:2}
\end{align}
Here, we define $z=\exp(- i  2\pi  q_y/b_2)$ with $b_2$ being the length of $\mathbf{b}_2$. And by definition, $n$ is an integer.

This proves that as a function of $z$, each element of the $\mathcal{C}$ matrix is a Laurent polynomial (i.e. $\sum_{n} c_n z^n$ with the power $n$ being an integer, 
which is not necessarily positive). More importantly, this Laurent polynomial must have a finite length, i.e. the sum contains only finite number of terms, instead of an infinitely 
long power series.
\begin{align}
(\mathcal{C})_{m,(\alpha)i}=\sum_{n=M_{\textrm{min}}}^{M_{\textrm{max}}} c_n z^n
%c\; z^{N_0} \;   \Pi_{i=1}^{M_0} (z-z_i)
%\frac{c_0+c_1 z +c_2 z^2+\ldots c_{M_0}z^{M_0}}{z^{N_0}}
\label{app:eq:poly:C:element}
\end{align}
with $c_n$ be some complex coefficients and the integer power $n$ must stay in a finite window $M_{\textrm{min}}\le n \le M_{\textrm{max}}$.
Here $M_{\textrm{min}}$ and $M_{\textrm{min}}$  are finite integers, and they could be positive, negative or zero.
This is because in the sum of Eq.~\eqref{app:eq:cmatrix:kspace}, each $\mathbf{t}$ implies a cross term in the elastic energy between two nodes whose unit cells are separated by  distance $\mathbf{t}$. For a local theory with short-distance interactions, this $\mathbf{t}$ cannot be large, and thus it sets a cap to the absolute value of the integer 
power $n$. 

Because the determinant is the sum of products of matrix elements, if each matrix element is a finite Laurent polynomial [Eq.~\eqref{app:eq:poly:C:element}], 
the determinant must also be a finite Laurent polynomial
\begin{align}
\det{\mathcal{C}}(z)=\sum_{n=N_{\textrm{min}}}^{N_{\textrm{max}}} c_n z^n
%=\frac{c_0+c_1 z +c_2 z^2+\ldots c_{M}z^{M}}{z^{N}}
\label{app:eq:poly:det:C}
\end{align}
The $c_n$ here is a set of coefficients different from the ones in Eq.~\eqref{app:eq:poly:C:element}. And the integer power $n$ is between 
$N_{\textrm{min}}$ and $N_{\textrm{max}}$.
Here, we used the fact that the $\mathcal{C}$ has a finite dimension, which is true for discrete Maxwell lattices where the dimension of the matrix is $N_d\times N_d$ 
with $N_d$ being the number of the degrees of freedom per cell. 

This $\det{\mathcal{C}}(z)$ can be rewritten as
\begin{align}
\det{\mathcal{C}}(z)=c \; z^{N} \;\Pi_{i=1}^{M}(z-z_i)
\label{app:eq:det:c:in:solution}
\end{align}
Here $\{z_i\}$ is a set of \emph{nonzero} complex constants, at which $\det{\mathcal{C}}(z_i)=0$.  $c=c_{N_{\textrm{max}}}$ is a complex constant, and the two integers 
$N=N_{\textrm{min}}$ and $M=N_{\textrm{max}}-N_{\textrm{min}}$ ($M\ge 0$ but $N$ cant be positive, negative or zero).

As shown in Sec.~\ref{app:sub:sec:zeros}, each $z_i$ corresponds to one zero modes with $\det{\mathcal{C}}(z_i)=0$. It is worthwhile to mention that in addition to these nonzero solutions ($z_i$), 
$\det{\mathcal{C}}(z)=0$ may have one additional solution at $z=0$ (if $N>0$) or $z=\infty$ (if $M+N<0$). As will be shown below,
they don't correspond to physical zero modes, and their contributions makes the topological polarization gauge dependent..

\subsection{Topological index and gauge choice}
\label{app:sub:sec:lattices:index}
As mentioned in Sec.~\ref{sec:kspace}, by choosing a different gauge, i.e., a different unit cell convention, the elastic energy associated with a unit cell 
$E_{R}$ [Eqs.~\eqref{app:eq:E:as:ER} and \eqref{app:eq:square:form:ER}] will change accordingly, which changes the $\mathcal{C}$ matrix
[Eq.~\eqref{app:eq:cmatrix:kspace}]. This change of $\mathcal{C}$ can be represented by 
multiplying a diagonal matrix $\mathcal{D}_1$ from the left side of $\mathcal{C}$, i.e., $\mathcal{C} \to \mathcal{D}_1\mathcal{C}$, where the 
diagonal elements of this diagonal matrix $\mathcal{D}_1$ are $z$ to some integer powers.
In addition, because some nodes will be assigned to a different cell as we change the unit cell convention, the effect also changes the 
associated phase factor in $\mathcal{C}$, which corresponds to multiplying an diagonal matrix $\mathcal{D}_2$ from the right side, i.e.
$\mathcal{C} \to \mathcal{C} \mathcal{D}_2$. Same as $\mathcal{D}_1$, the diagonal elements of $\mathcal{D}_2$ are also $z$ to some integer powers.

As a result, changing unit cell convention changes $\det \mathcal{C}$ into $\det \mathcal{D}_1\det \mathcal{C}\det \mathcal{D}_2$, i.e
\begin{align}
\det \mathcal{C} \to \det \mathcal{D}_1\det \mathcal{C}\det \mathcal{D}_2 =z^{\Delta N} \det \mathcal{C},
\end{align}
where $\Delta N$ is some integer. This immediately implies that in Eq.~\eqref{app:eq:det:c:in:solution}, the integer power $N$ is gauge 
dependent, where the product part $\Pi_{i=1}^{M}(z-z_i)$ is gauge invariant. 
Depending on how we choose a unit cell, the value of $N$ can be shifted by an arbitrary integer, but the product part always remain. 
As will be shown below, the product part is why the topological index can detect edge modes, while the $z^N$ part is the reason why 
the topological polarization is gauge dependent.

In a Maxwell lattice, the topological index, i.e., each component of topological polarization, is defined as
\begin{align}
n_i=\frac{1}{2\pi i}\oint d z \tr \left(\mathcal{C}^{-1} \partial_{z}\mathcal{C}\right)
=\frac{1}{2\pi i}\oint d z \tr \left(\partial_{z} \ln \mathcal{C}\right)
=\frac{1}{2\pi i}\oint d z \;\partial_{z} \ln \det\mathcal{C} ,
\end{align}
where the contour is the unit circle of the complex $z$ plane.
Using Eq.~\eqref{app:eq:det:c:in:solution}, it is easy to verity that 
\begin{align}
\partial_z \ln \det \mathcal{C} (z)=\sum_{i=1}^{M} \frac{1}{z-z_i} -\frac{N}{z}+ c ,
\label{app:eq:poles:and:residue}
\end{align}
and thus
\begin{align}
n_i=\frac{1}{2\pi i}\oint d z \left(\sum_{i=1}^{M} \frac{1}{z-z_i}  \right)-\frac{1}{2\pi i}\oint d z \frac{N}{z} .
\label{app:eq:poles:and:residue:integral}
\end{align}
Here, we emphasize again that $z_i\ne 0$. As discussed above, the first integral is gauge invariant, but the second integral is gauge dependent, because the value of the integer $N$ changes if a different unit-cell definition
is adopted. 

In Maxwell lattices, the integral contour is usually set as the unit circle on the complex $z$ plane, and this integral can be easily computed utilizing the Cauchy integral theorem.  
\begin{align}
n_i= N_B+ N ,
\end{align}
where $N_B$ is the number of $z_i$'s inside the unit circle. Because each $z_i$ corresponds to a zero mode, as shown above, $N_B$ counts
the number of zero modes whose $z_i$'s are located inside the unit circle (aside from at the origin). 
By definition, $z_i$ inside the unit circle means that the imaginary part of 
$q_y$ is positive, and thus according to Eq.~\eqref{app:eq:wave:form}, such a zero mode is localized at the bottom edge of a strip. In other words, $N_B$ gives
the number of zero modes for the bottom edge. This is the key reason why $n_i$ is sensitive to the number of edge modes for each edge.

However, in addition to $N_B$, $n_i$ also contains a gauge-dependent contribution $N$, whose value is sensitive to the unit-cell convention. This gauge-dependent $N$
is the reason why the value of $n_i$ changes if a different unit cell is adopted. As shown in the loop integral above, this gauge-dependent part
comes from the pole of $\partial_{z} \ln \det\mathcal{C}$ at $z=0$. Thus, to obtain an gauge independent topological index, this pole should be removed from the integral,
which can be achieved utilizing the contour shown in Fig.~3(c) in the main text (i.e., the boundary of the region B).

If we reverse the direction of the unit-circle contour and then applies the Cauchy integral theorem, then $-n_i$ should equals to the total residue summed over all poles
outside the unit circle. For each $z_i$ with $|z_i|>1$, it gives a pole with residue $1$, and thus contributes $1$ to the integral. Because $|z_i|>1$ implies $q''_y<0$, these
are edge modes on the top edge, i.e., each edge mode on the top edge gives $1$ to the integral.  In addition,
$\partial_z \ln \det \mathcal{C} (z)$ also have a pole at $z=\infty$ and its residue is $-(N+M)$. This is because at $z=\infty$, $1/z$ gives a residue $-1$, and in addition each $1/(z-z_i)$
also has a residue $-1$ at $z=\infty$. Thus, from Eq.~\eqref{app:eq:poles:and:residue:integral}, the total residue at $z=\infty$ is $N-M$. After adding all the poles ($z=z_i$ with 
$|z_i|>1$
and $z=\infty$), we get
\begin{align}
-n_i= N_T - (N+M)
\end{align}
where the first term comes from zero-energy modes at the top edge, and the second term is from the pole at $z=\infty$. Again, the first term here
is gauge invariant, but the second term is gauge dependent.  It is worthwhile to mention that this result agrees with $n_i= N_B+N$ that we obtained above. This can be easily checked
by noticing that $M=N_B+N_T$. Here, we can avoid the gauge-dependent contribution from the pole at $z=\infty$ by using the contour shown in Fig.~3(c) in the main text 
(i.e. the boundary of the region T). Once $z=\infty$ is removed, the gauge-dependent second term ($N-M$) disappear and the topological index directly gives $N_T$.

In kagome lattices, one frequently used gauge has $N=-1$. In the non-polarized phase, where each edge has one zero mode per unit cell ($N_B=N_T=1$), 
this leads to $n_i=N_B+N=0$.
For the polarized phase, both two zero modes are located on the bottom edge ($N_B=2$ and $N_T=0$), and thus $n_i=N_B+N=1$. By keeping 
$z=0$ and $z=\infty$ out of the integral contour, the topological index directly gives $N_B$ and $N_T$ without the gauge dependent part $N$.

In summary, in this section we showed that $\tr\mathcal{C}^{-1} \partial_{z}\mathcal{C}$ is analytic for the entire complex plane, except for $M+2$ poles.
Among these poles, $M$ of them comes from zero modes, each of which has residue $1$ and is gauge invariant. In addition to physical zero modes, 
there are two additional poles located at $z=0$ and $z=\infty$. As will be shown in the next section, these two poles are not due to zero modes. Instead, they are
required by analytic continuation (as we extend the definition of $\det \mathcal{C}$ from real $q_y$ to the entire complex $z$ plane), and their contribution
to the topological index is gauge dependent. By choosing a new set of contours, their contributions can be removed, which leads to a gauge independent topological index.

Finally, it is worthwhile to mention that in the calculation above, we assumed that $\det \mathcal{C} (z)\ne 0$ on the unit circle ($|z|=1$). Because the unit circle
corresponds to real $q_y$, this means that the bulk phonon modes (at this fixed nonzero $q_x$) are fully gapped without bulk zero mode. 
If this is not the case, the topological index becomes ill-defined and the system (at this $q_x$) is at the phase boundary between two different topological phases 
(i.e., a topological phase transition~\cite{Rocklin2017} or mechanical Weyl points~\cite{Rocklin2016}).

\subsection{Dangling bonds and analytic continuation}
\label{app:sub:sec:lattices:dangling}
In this section, we discuss two subjects (a) edge modes from dangling bonds and (b) why analytic continuation necessarily leads to poles that are not related with edge modes.

Here, we use the terminology ``dangling bonds'' broadly to refer to all zero-energy edge modes with zero penetration depth, i.e., the imaginary part of $q_y$ is infinity. 
Due to its local nature (i.e. zero penetration depth), these zero modes are associated  local degrees of freedom at the edge, instead of a bulk origin. Thus, bulk quantities
like bulk topological indices are not expected to capture this physics. In terms of $z$, these ``dangling bonds'' corresponds to $z=0$ (for $q''_y=+\infty$) and
$z=\infty$ (for $q''_y=-\infty$). Thus, because bulk physics is not expected to capture zero modes at  $z=0$ and $z=\infty$,
contributions from poles at these to points are not expected to carry correct information about the edge, and this is the reason why they are
gauge dependent.  Furthermore, when a Maxwell lattice is cut at different depths at an edge (e.g., shifting the cut up or down for a horizontal edge), the number of edge zero modes that have a topological origin (i.e., due to bulk polarization) is invariant, and only the number of zero modes due to dangling bonds changes.

Here, we demonstrate that due to analytic continuation, the poles of $\tr \left(\mathcal{C}^{-1} \partial_{z}\mathcal{C}\right)$ cannot all correspond to physical edge modes.
As mentioned above, each zero modes corresponds to a pole with residue $1$ in $\tr \left(\mathcal{C}^{-1} \partial_{z}\mathcal{C}\right)$, which is the reason why
the topological index is sensitive to zero mode. However, analytic continuation requires this function to have additional poles, beyond those who come from zero modes.
This is because according to the Cauchy integral theorem, the total residue from all poles in the entire complex plane must be zero (we showed above that $\mathcal{C}$ 
has no other singularity beyond poles for the entire complex plane). Thus, there must be some poles with negative residue to counter balance the poles from edge modes, which
is the only way to keep total residue zero. Where can these additional poles be located? As shown above, $z=0$ and $z=\infty$ corresponds to 
local degrees of freedom at the boundary, and thus for a bulk topological index, they are the perfect place for these extra poles.

\subsection{Continuum theory}
\label{app:sub:sec:continuum}
In the continuum theory, same as shown in Sec.~\ref{app:sub:sec:zeros}, zero modes are also associated with zeros in $\det \mathcal{C}$.
Thus, in terms of the complex variable $z$ [i.e., Fig~2.(b) in the main text], again we can search for edge modes by finding zeros of $\det \mathcal{C}(z)$.

Different from  lattice models, here, the continuum theory is only expected to hold at long distance (small $q$), thus in the complex $z$ plane, the theory only applies
to a domain near $z=1$ (i.e., $q_y=0$).  In the vicinity of $z=1$, $\det \mathcal{C}(z)$ should be an analytic function and thus can be written as a power law expansion, 
\begin{align}
\det \mathcal{C}(z) = \sum_{n\ge 0} c_n (z-1)^n.
\end{align}
In contrast to lattices, this sum contains no negative power. In addition, this power-law expansion is only valid in a domain around $z=1$. Outside this domain,  
this function in general is not meromorphic for the entire complex plane, and essential singularities that cannot be represented as poles or branch cuts may arise as 
we move far away from $z=1$. This is the mathematical reason why a cut off needs to be introduced, which keeps us inside this domain.

Because we always stay inside the convergence domain of the power law expansion, similar to the lattice models discussed above, one can easily prove that 
here, if $\det \mathcal{C}(z_i)=0$ and $z_i$ is inside the convergence domain, $\tr \left(\mathcal{C}^{-1} \partial_{z}\mathcal{C}\right)$ shall have a pole at $z=z_i$
with residue $1$. Thus, the same integral shown above counts the number of zero modes inside the integral contour.

Because our power expansion here has no negative power and because our integral contour is alway inside the convergence domain, there is no unphysical pole, i.e., poles inside
our contour   has a one-to-one correspondence with physical zero modes and their contribution to the integral is gauge invariant. This is why this topological index 
is gauge independent and directly measures the number of zero modes.

Finally, we emphasize that the integral contour utilized in the continuum theory only covers part of the unit circle. Thus, for a Maxwell lattice,
it is possible that there might exist zero modes whose $z_i$'s are outside the integral contour of the continuum theory. For those zero modes, because their $z_i$'s
is far away from $1$, i.e. $q_y$ far away from $0$, they involve short-distance physics, which the continuum theory is not expected to capture. %But for all long-distance
%physics (i.e. zero modes with small $q_y$), the continuum theory will not miss them.

\section{Beyond Maxwell lattices}
In this section, we present the variational method discussed in the main text. For systems not far from the Maxwell point, we can first treat the system as a Maxwell medium by ignoring terms associated with small eigenvalues, as shown in the main text.
Within this zeroth order approximation, the continuum theory is at the Maxwell point, and we can compute the topological index and obtain the decay rate of the edge modes. 
Afterwards, we can use this zeroth order results as input and compute the energy expectation value of this mode.

Similar to the ground state of a quantum system, at a fixed wave vector, the deformation configuration that minimizes the elastic energy is the normal mode with the
lowest eigen-energy. Instead of performing a full minimization, here we use the idea of the variational principle. 
Instead of searching all possible real space configurations,  we limit our deformation to the following functional form
\begin{align}
\mathbf{u}=\mathbf{A}\exp(i \bar{q}_x x+i q'_y y-q''_y y).
\end{align}
Here we fix one component of the wave vector along the edge direction ($q_x=\bar{q}_x$), and the amplitude $\mathbf{A}$ as well as 
$q'_y$ and $q''_y$ are variational parameters, which we will adjust in search for the lowest-energy mode. 
Such a minimization will produce an estimation of the normal mode with lowest eigen energy at $q_x=\bar{q}_x$. If $q''_y=0$, it is a bulk mode. If $q''_y\ne 0$,
it is an edge mode.
In a rigid solid, the true lowest-energy mode (e.g. Rayleigh waves) don't take this form. Instead, they are often composed of linear combination of multiple  
exponentials, each of which contains a different set of $q'_y$ and $q''_y$. However, in our leading order theory at the Maxwell point, the mode with lowest energy 
indeed take this form, and thus as long as a system is not far from the ideal Maxwell limit, this waveform provides a good approximation.

It should be emphasized that such a constrained minimization is not expected to produce the exact answer. Instead, it always over estimate the energy for the lowest energy mode.
However, if the initial guess we made are reasonably close to the real minimum, this approach gives a good upper bound for energy. In our calculation here,
we will further simplify the problem by setting $q'_y$  and $q''_y$ to be what we obtained from the zeroth order theory. As mentioned early on, at the zeroth order,
we can use the topological theory to obtain edge modes, as well as their $q'_y$  and $q''_y$, which are the values that we will use here. The idea here
is that we are using the profile revealed by the zeroth-order topological theory to guide our variational calculation. As long as the system is not far from the ideal Maxwell point, 
this zeroth order theory should produce reliable estimation about the edge modes.

By plugin this zeroth order profile into the elastic energy Eq.~\eqref{app:eq:square:form:ER}, it is easy to realize that the energy expectation value is
\begin{align}
\langle E_R \rangle=\frac{1}{2}\frac{ \mathbf{A}^\dagger [\mathcal{C}(\bar{q}_x, q'_y+i q''_y)]^\dagger\mathcal{C}(\bar{q}_x, q'_y+i q''_y)  \mathbf{A}}{\mathbf{A}^\dagger\mathbf{A}} ,
\end{align}
where the $\mathcal{C}$ matrix is defined in Eq.~\eqref{app:eq:cmatrix:kspace}, and the amplitude for each node $\mathbf{A}$ is our variational parameter. 
Because the matrix $[\mathcal{C}(\bar{q}_x, q'_y+i q''_y)]^\dagger\mathcal{C}(\bar{q}_x, q'_y+i q''_y)$ is unitary, minimizing $\langle E_R \rangle$ here corresponds
to finding the lowest eigenmode of this matrix. Thus, within this variational method, the energy of the lowest-energy mode in our system is the lowest eigenvalue of
$[\mathcal{C}(\bar{q}_x, q'_y+i q''_y)]^\dagger\mathcal{C}(\bar{q}_x, q'_y+i q''_y)$, and its square root gives a good estimation about the frequency of the mode,  
which is what we plotted in Fig.~3.
Because the zeroth order theory gives  two edge modes for each $q_x$ (with different $q'_y$ and $q''_y$ in general), the procedure discussed above are performed 
for each of them, which gives the two edge-mode branches in Fig. 3. 
As can be seen in Fig.~3, in comparison with full scale numerical simulations,  these two curves slightly overestimate the mode frequency, 
which is expected due to the variational nature of our method.

It must be emphasized that here we are searching for the lowest eigenvalue of $[\mathcal{C}(q_x, q'_y+i q''_y)]^\dagger\mathcal{C}(q_x, q'_y+i q''_y)$. This 
should not be confused with the lowest eigenvalue of the dynamic matrix $\mathbf{D}(q_x, q'_y+i q''_y)$. 
For bulk modes (i.e., real $q_x$ and $q_y$), $\mathbf{D}(q_x, q_y)= [\mathbf{C}(q_x, q_y)]^\dagger\mathbf{C}(q_x, q_y)$
as have been shown above, but it is not the case for edge modes with complex $q_y$.
This is because  in reality, the dynamic matrix is defined as 
\begin{align}
\mathbf{D}(q_x, q_y)= [\mathbf{C}(-q_x, -q_y)]^T\mathbf{C}(q_x, q_y) ,
\label{app:eq:dmatrix:definition}
\end{align}
instead of $\mathcal{C}^\dagger\mathcal{C}$. For real $q_x$ and $q_y$, $[\mathbf{C}(-q_x, -q_y)]^T$ is identical to $[\mathbf{C}(q_x, q_y)]^\dagger$. However, because Eq.~\eqref{app:eq:dmatrix:definition}
is what is really utilized in the definition of the dynamic matrix, as we generalize $q_y$ into complex values (for the edge modes), the dynamic matrix becomes
\begin{align}
\mathbf{D}(q_x, q'_y+iq''_y)= [\mathbf{C}(-q_x, -q'_y-i q''_y)]^T\mathbf{C}(q_x, q'_y+iq''_y),
\end{align}
which  differs sharply from the matrix $[\mathcal{C}(q_x, q'_y+i q''_y)]^\dagger\mathcal{C}(q_x, q'_y+i q''_y)$ that we used above, Most importantly, in contrast to our
hermitian matrix $[\mathcal{C}(q_x, q'_y+i q''_y)]^\dagger\mathcal{C}(q_x, q'_y+i q''_y)$, the matrix $\mathbf{D}(q_x, q'_y+iq''_y)$ is not hermitian with a complex $q_y$, 
and thus its eigenvalues have no direct physical meanings. This is the key reason why edge-mode information is not obtained from the dynamic matrix.

\section{Lattice models}
In this section, we present details about the lattice modes that we used for Figs~2 and 3 in the main text and the coefficients of their continuum theory.

The lattice we use is shown in Fig.~\ref{fig:latt}, which is a topological kagome lattice with point-like sites connected by central-force bonds (balls connected by ideal springs).  Sides of red triangles are of lengths $(1,1,1)$ and sides of the blue triangles of at lengths $(1,1/\sqrt{3},1/\sqrt{3})$.  The angle $\theta$ between the two triangles is $\theta=5\pi/12$.  Each unit cell contains 3 sites (e.g., the 3 vertices of the red triangles), and the lattice primitive vectors are $a_1=(\sqrt{2 + \sqrt{2 + \sqrt{3}}}, 0)$, 
$a_2=(-\sqrt{ (2 + \sqrt{2})/3}, \sqrt{(2 + \sqrt{2 + \sqrt{3}})/3})$, as shown in the figure.  With nearest neighbor bonds only, this lattice has $z=4$ and is Maxwell.  When next-nearest neighbor bonds are included, each site has an extra coordination of 4 above the Maxwell point.

For Fig.~2 in the main text, we used this lattice with only nearest neighbor bonds, and we took all spring constants to be 1. The topological polarization is $R_T = a_2$, implying that the top edge has two zero modes per edge unit cell, while the bottom edge is rigid with no zero mode.

\begin{figure}[t]
	\centering
	\includegraphics[width=.4\columnwidth]{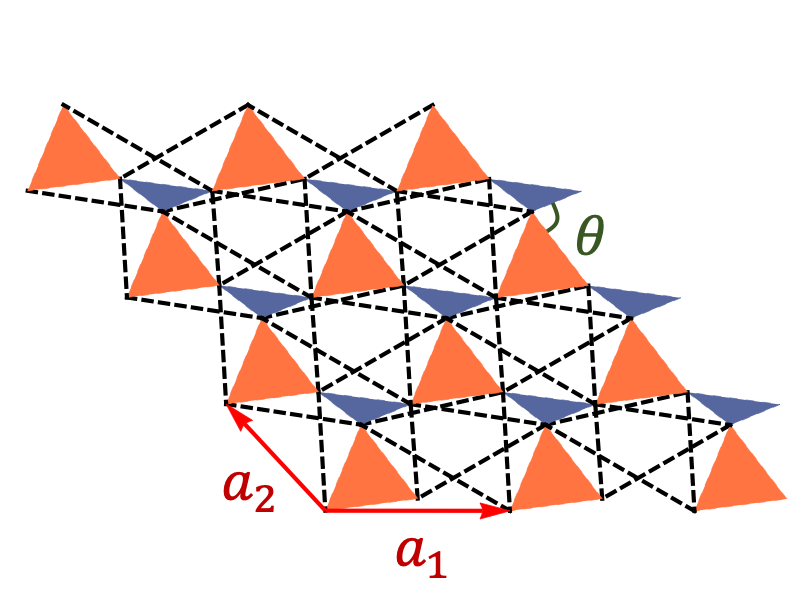}
	\caption{The topological kagome lattice used for Figs.~2 and 3 in the main text. 
	Each edge of a triangle represent a  nearest-neighbor spring with spring constant $1$. The dashed lines represent next-nearest-neighbor 
	springs. Without next-nearest-neighbor springs, the lattice is at the Maxwell point with two zero modes per wave vector $q_x$ at the top edge, while the bottom edge is rigid without zero modes (Fig.~2 in the main text).
	If next-nearest-neighbor springs are introduced, the lattice is over-constrained and a lattice topological index cannot be defined within the existing 
	theoretical framework for discrete Maxwell lattices, while our continuum theory allows a topological description and the energy of the edge modes are shown in Fig. 3 in the main text.}
	\label{fig:latt}
\end{figure}

%%%%%%%%%%%
%geometry of the lattice needs to be added
%Needs the length of a_1 and a_2 (a_1 // to edge)
%%%%%%%%%%%%

For Fig.~3 in the main text, we used a same  lattice but next-nearest-neighbor springs are introduced, which makes the system over constrained. The spring constant of the 
next-nearest-neighbor springs are set to $1/1000$, which is much smaller than the spring constant of the nearest-neighbor springs ($1$). 
Because the next-nearest-neighbor springs are weak, this system is not far from the ideal Maxwell point, and thus the top edge contains two soft modes, whose frequencies
are no longer exactly zero, while the bottom edge remains rigid.  

Following the procedure shown above, we computed the elastic constants for the continuum theory of these two models. For the Maxwell lattice (Fig.~2),
we used two different continuum model, one with only two acoustic bands and the other contains also a soft optical band.  Both of them yield topological index agreeing with the discrete lattice theory. 

For the first model where we keep two acoustic bands [Fig.~2(a)], the elastic energy is
\begin{align}
E=\lambda_1 \xi_1^2 +\lambda_2 \xi_2^2 ,
\end{align}
where $\lambda_1=0.731617$ and $\lambda_2=0.130026$, and
\begin{align}
\xi_1=&[0.889562 \partial_x u_x-0.285524 (\partial_y u_x+\partial_x u_y)+0.213618 \partial_y u_y]
\nonumber\\
&+ [(-0.168009 \partial_x^2 + 0.302434 \partial_x \partial_y + 0.17858 \partial_y^2)u_x
+(0.0945483 \partial_x^2  + 0.204106 \partial_x\partial_y - 0.0537692 \partial_y^2)u_y] ,
\\
\xi_2=&[0.320964 \partial_x u_x+0.155366 (\partial_y u_x+\partial_x u_y)-0.921252 \partial_y u_y]
\nonumber\\
&+ [(0.266878 \partial_x^2 + 1.34449 \partial_x \partial_y + 0.247074 \partial_y^2)u_x 
+(0.750664 \partial_x^2  - 0.0806679 \partial_x\partial_y + 0.127889 \partial_y^2)u_y] .
\end{align}
At small $q_x$, the $q_y$ of the two edge modes have the following asymptotical form
\begin{align}
q_y=&0.140684 q_x a_1 +  0.231601 (q_x a_1)^2  i ,
\\
q_y=&1.80784 q_x a_1 +  3.82382 (q_x a_1)^2 i ,
\end{align}
and the imaginary part of $q_y$ is plotted as the dotted lines in Fig.~2(a).

For the second model where we also include the lowest optical band [Fig.~2(b)], the elastic energy is
\begin{align}
E=\lambda_1 \xi_1^2 +\lambda_2 \xi_2^2 +\lambda_3 \xi_3^2 ,
\end{align}
where $\lambda_1=0.922319$, $\lambda_2=0.476694$ and 
$\lambda_3= 0.126092$ and
\begin{align}
%(0. + 0. I) - (0. + 0.901407 I) q1 - 0.160218 q1^2 - (0. + 0.0915319 I) q2 + 0.244429 q1 q2 +  0.0791255 q2^2
%(0. + 0.101537 I) q1 + 0.0705272 q1^2 - (0. + 0.211443 I) q2 +  0.119816 q1 q2 + 0.00669479 q2^2
%(-0.352234 + 0. I) + (0. + 0.0272962 I) q1 - (0. + 0.105238 I) q2
\xi_1=&[0.901407 \partial_x u_x+0.0915319 \partial_y u_x
-0.101537 \partial_x u_y+0.211443 \partial_y u_y
+0.352234 \phi]
\nonumber\\
&+ [(0.160218 \partial_x^2 - 0.244429 \partial_x \partial_y - 0.0791255 \partial_y^2)u_x
+(-0.0705272 \partial_x^2  -0.119816 \partial_x\partial_y - 0.00669479 \partial_y^2)u_y
\nonumber\\
&\;\;\;\;\;\;\;\;-0.0272962  \partial_x\phi +  0.105238 \partial_y\phi],
\\
%(0. + 0.226866 I) q1 + 0.249133 q1^2 - (0. + 0.690368 I) q2 +  0.241838 q1 q2 - 0.0151846 q2^2
%-(0. + 0.380851 I) q1 + 0.0089196 q1^2 + (0. + 0.0894918 I) q2 - 0.0835355 q1 q2 + 0.172733 q2^2
%(-0.564684 + 0. I) - (0. + 0.170975 I) q1 - (0. + 0.421068 I) q2
\xi_2=&[0.226866 \partial_x u_x-0.690368 \partial_y u_x
-0.380851\partial_x u_y + 0.0894918\partial_y u_y
-0.564684\phi]
\nonumber\\
&+ [(0.249133 \partial_x^2 + 0.241838 \partial_x \partial_y - 0.0151846 \partial_y^2)u_x
+(0.0089196 \partial_x^2  - 0.0835355 \partial_x\partial_y +  0.172733 \partial_y^2)u_y
\nonumber\\
&\;\;\;\;\;\;\;\;+0.170975\partial_x\phi+ 0.421068 \partial_y\phi] ,
\\
%(0. + 0.282761 I) q1 + 0.982133 q1^2 + (0. + 0.0258273 I) q2 -  0.0541133 q1 q2 + 0.342373 q2^2
%(0. + 0.0997509 I) q1 - 0.391609 q1^2 - (0. + 0.944055 I) q2 +  0.432983 q1 q2 - 0.0747353 q2^2
%(-0.134866 + 0. I) - (0. + 1.47841 I) q1 - (0. + 0.192398 I)
\xi_3=&[0.282761 \partial_x u_x+0.0258273 \partial_y u_x
+0.0997509\partial_x u_y -0.944055 \partial_y u_y
-0.134866\phi]
\nonumber\\
&+ [(0.982133 \partial_x^2 -0.0541133 \partial_x \partial_y + 0.342373 \partial_y^2)u_x
+(-0.391609 \partial_x^2+0.432983 \partial_x\partial_y -0.0747353 \partial_y^2)u_y
\nonumber\\
&\;\;\;\;\;\;\;\;+1.47841\partial_x\phi-0.192398 \partial_y\phi] ,
\end{align}
Here $\phi$ is a softest optical phonon mode. In contrast to the elastic theory with only acoustic modes, here, the $\partial_y u_x$ term and the $\partial_x u_y$ term are
not symmetric and they have different coefficients. This is expected because we included an optical phonon mode. Without optical modes, $\partial_y u_x$ and 
$\partial_x u_y$ must have the same coefficient, because the antisymmetric combination  $\partial_y u_x -\partial_x u_y$ corresponds to rigid body rotation
and thus shall not arise in the elastic energy. In our elastic energy above, this is still true and the antisymmetric part $\partial_y u_x -\partial_x u_y$ can be absorbed
into a redefinition of $\phi$, if we replace $\phi$ with $\widetilde{\phi}=\phi+0.274063 (\partial_y u_x-\partial_x u_y)$. After this redefinition,  the $\partial_y u_x$ term and 
the $\partial_x u_y$ term will become symmetric.

At small $q_x$, the $q_y$ of the two edge modes have the same asymptotical form as shown in the two bands model
\begin{align}
q_y=&0.140684 q_x a_1 +  0.231601 (q_x a_1)^2  i ,
\\
q_y=&1.80784 q_x a_1 +  3.82382 (q_x a_1)^2 i ,
\end{align}
and the imaginary part of $q_y$ is plotted as the dotted lines in Fig.~2(b).
The agreement of the two-band and three-band model at the small $q_x$ limit serves as a consistency check for our calculations.

After the next-nearest-neighbor springs are introduced, the two-band continuum theory becomes
\begin{align}
E=\lambda_1 \xi_1^2 +\lambda_2 \xi_2^2 +\lambda_3 \xi_3^2 ,
\end{align}
with $\lambda_1=0.734928$, $\lambda_2=0.131478$, and $\lambda_3=0.00869398$. Because the $\lambda_3$ is very small, we will drop it as the zero order approximation,
i.e., $r_M=\lambda_3/\lambda_2=0.066125$. This zeroth order theory is at the Maxwell point and thus we can perform topological analysis as shown above. And afterwards, we use the
variational method to compute the energy of these topological modes. The square root of the energy, which is the frequency, is shown in Fig.~3(a).

In Fig.~3(b), we included an optical mode and the continuum elastic energy becomes
\begin{align}
E=\lambda_1 \xi_1^2 +\lambda_2 \xi_2^2 +\lambda_3 \xi_3^2 + \lambda_4 \xi_4^2 ,
\end{align}
where $\lambda_1=0.925252$, $\lambda_2=0.478491$, $\lambda_3=0.1272$, and $\lambda_4=0.00470203$
and $r_M=\lambda_4/\lambda_3=0.0369656$. Because  $r_M$ is small, we can ignore $\lambda_4$ as the leading order theory, which is at the Maxwell point.
The topological indices and zero modes are obtained from this leading order theory, whose energy is then computed using the variational method. Its square root,
i.e., the frequency, is plotted in Fig.~3(b).

\end{widetext}

\bibliography{isostaticity}

\end{document}